\documentclass[12pt]{iopart}

\usepackage[usenames,dvipsnames]{xcolor}
\definecolor{articleColor}{cmyk}{0.3 , 0.3  , 0   , 0.09} 
\usepackage[utf8]{inputenc}
\usepackage[normalem]{ulem} 
\usepackage{amsthm, amsfonts, amssymb}
\usepackage{siunitx}
\usepackage{array}
\usepackage{times}
\usepackage{latexsym}
\usepackage{hyperref}
\hypersetup{backref,  
colorlinks=true,
linkcolor=red,
filecolor=red,
citecolor=blue}
\usepackage{epsfig}
\usepackage{bm}

\begin{document}

\title{Initial data for black hole--neutron star binaries, with rotating stars}

\newcommand{\AEI}{{Max Planck Institute for Gravitational Physics
(Albert Einstein Institute), Am M\"uhlenberg 1, Potsdam-Golm, 14476, Germany}} %
\newcommand{\Caltech}{{TAPIR,
    Walter Burke Institute for Theoretical Physics 350-17,
    California Institute of Technology, Pasadena, CA 91125, USA}}
\newcommand{\CITA}{{Canadian Institute for Theoretical
    Astrophysics, University of Toronto, 60~St.~George Street,
    Toronto, Ontario M5S 3H8, Canada}} %
\newcommand{\CIFAR}{{Canadian Institute for Advanced
    Research, 180 Dundas St.~West, Toronto, ON M5G 1Z8, Canada}} %
\newcommand{\Cornell}{{Cornell Center for Astrophysics and Planetary
  Science, Cornell University, Ithaca, New York 14853, USA}}
\newcommand{\DAA}{{Department of Astronomy and
    Astrophysics, 50 St.\ George Street, University of Toronto,
    Toronto, ON M5S 3H4, Canada}}
\newcommand{\LBL}{{Lawrence Berkeley National Laboratory,
    1 Cyclotron Rd, Berkeley, CA 94720, USA}}
\newcommand{\WSU}{{
Department of Physics \& Astronomy, Washington State University, Pullman, Washington 99164, USA}}

\author{Nick Tacik$^1$$^2$,          
        Francois Foucart$^3$,        
        Harald P. Pfeiffer$^1$$^4$,  
        Curran Muhlberger$^5$,       
        Lawrence E. Kidder$^5$,      
        Mark A. Scheel$^6$,          
        B\'{e}la Szil\'{a}gyi$^6$}   
\address{$^1$ \CITA}
\address{$^2$ \DAA}
\address{$^3$ \LBL}
\address{$^4$ \CIFAR}
\address{$^5$ \Cornell}
\address{$^6$ \Caltech}

\date{\today}

\begin{abstract}
  The coalescence of a neutron star with a black hole is a primary
  science target of ground-based gravitational wave detectors.
  Constraining or measuring the neutron star spin directly from
  gravitational wave observations requires knowledge of the dependence of
  the emission properties of these systems on the neutron star spin.
  This paper lays foundations for this task, by developing a numerical
  method to construct initial data for black hole--neutron star
  binaries with arbitrary spin on the neutron star. We demonstrate
  the robustness of the code by constructing initial-data sets in
  large regions of the parameter space. In addition to varying the
  neutron star spin-magnitude and spin-direction, we also explore
  neutron star compactness, mass-ratio, black hole spin, and black
  hole spin-direction. Specifically, we are able to construct initial
  data sets with neutron stars spinning near centrifugal break-up, and
  with black hole spins as large as $S_{\rm BH}/M_{\rm BH}^2=0.99$.
\end{abstract}


\submitto{\CQG}

\section{Introduction}
\label{sec:Introduction}

The spectacular detection of merging binary black holes by Advanced
LIGO~\cite{LIGOVirgo2016a,Abbott:2016nmj} marks the beginning of the
era of gravitational wave astronomy. With binary black holes detected
through gravitational waves, and binary neutron stars known from radio
observations~\cite{Hulse:1975uf}, mixed black-hole - neutron star
(BH-NS) binaries are now the only compact object binary whose
existence has not yet been directly observed.

BH-NS systems are an important potential source of gravitational waves
for advanced ground-based detectors, with an expected event rate of
approximately ten per year~\cite{AbadieLSC:2010}, albeit with a large
uncertainity. In addition to gravitational waves, BH-NS mergers can
be an important source of electromagnetic radiation~\cite{Li:1998bw,Roberts2011,metzger:11,2013MNRAS.430.2121P,2013MNRAS.430.2585R,2014ApJ...780...31T}
 and give further clues to
the violent processes that occur during the merger. If a massive disk
is left from the merger, for instance, it could lead to a
short-duration gamma ray burst (SGRB) and material ejected during the
merger could radiate a signal such as a "kilonova"~\cite{metzger:11}.

Direct numerical solutions are one of the primary means to explore
coalescing compact object binaries
(e.g.~\cite{baumgarteShapiroBook,2014ARA&A..52..661L,Pfeiffer:2012pc}).
Such simulations are important to accurately study both the
gravitational waves and electromagnetic emission produced by compact
object mergers. Fully general relativistic simulations of mixed BH-NS
binaries have been performed for about 10
years~\cite{Shibata:2006bs,Faber2005} investigating the importance of
mass-ratio~\cite{Foucart:2014nda,Foucart:2013psa,FoucartEtAl:2011},
black hole
spin~\cite{East:2011xa,Shibata:2006ks,Foucart:2013a,Foucart:2010eq,Kawaguchi:2015,Etienne:2008re},
eccentricity~\cite{East:2015yea,2012PhRvD..85l4009E,Stephens:2011as}, equations of state~\cite{Duez:2009yy,Kyutoku:2010zd,Kawaguchi:2015,Foucart:2013a},
magnetic
fields~\cite{Chawla:2010sw,Paschalidis2014,Kiuchi:2015qua,2012PhRvD..85f4029E,Etienne:2012te},
neutrino physics~\cite{Foucart:2015a}, disk
formation~\cite{Lovelace:2013vma,Shibata:2007zm,Pannarale:2015jia},
outflows~\cite{Deaton2013,Kyutoku:2013wxa} and electromagnetic
emission signatures~\cite{PaschalidisEtAl:2013,Kawaguchi:2016}.

The parameter space for BH-NS binary simulations is relatively
large. The mass ratio, $q$, NS compactness, $C$, and black hole spin, $\vec{\chi}$, have
been of particular interest in numerical simulations, because they
have the most profound impact of the evolution of the binary, and are the primary 
variables to control whether the neutron star tidally disrupts~\cite{Foucart2012}.
One aspect that has not been studied, however, is the effect of
neutron star spin. With the exceptions of~\cite{Shibata:2006bs,East:2015yea}, all
simulations to date use irrotational neutron stars in their BH-NS
binaries. For NS-NS binaries, in constrast, a significant number of
studies investigate spinning neutron
stars~\cite{Baumgarte:2009fw,Tichy:2011gw,East:2012zn,Tichy:2012rp,Bernuzzi:2013rza,Kastaun:2013mv,Tsatsin:2013jca,Dietrich:2015pxa,East:2015yea,Tsokaros:2015fea,Tacik:2015tja}. Since
no BH-NS binaries have been directly observed, the NS spins are, at
least observationally, unconstrained. A spinning neutron star will
affect the gravitational waveforms and cause the inspiral to proceed
more slowly (for spin-aligned NS). The spin can be important for
gravitational wave detection and can cause appreciable mismatch with
non-spinning templates, especially at lower BH-NS mass ratios
(\cite{Ajith:2011ec}). We also expect the spin to also affect the time of
NS disruption, as the stellar material will be less tightly bound to
the stellar surface.

Any evolution must start with initial data, and so in this paper, we
consider the construction of fully general-relativistic initial data
sets for BH-NS binaries with generic spin on the neutron star. We
combine the techniques of constructing BH-NS initial data without
NS-spin~\cite{FoucartEtAl:2008,Henriksson:2014tba} with the
rotating-NS formalism developed by Tichy~\cite{Tichy:2012rp} as
implemented in Tacik et al~\cite{Tacik:2015tja}. We show that this
approach, implemented in the Spectral Einstein Code {\tt
  SpEC}~\cite{SpECwebsite}, is robust and can construct BH-NS binaries
with NS spin magnitudes up to nearly rotational break-up
(dimensionless spin $\chi_{\rm NS}\sim 0.7$) and arbitrary rotation axis.
The code also successfully constructs binaries with mass-ratios
from 2 to 10, and with black hole spins $0\le \chi_{\rm BH}\le 0.99$.

The structure of this article is as follows: In
section~\ref{sec:BH-NSIDF}, we review the standard numerical
relativity initial data formalism, as well as the formalism developed
in \cite{Tichy:2011gw} to create binaries with spinning NS, and
discuss how this is extended to BH-NS systems. In
section~\ref{sec:BHNSNumMethods} we discuss the numerical methods used
by our initial data solver. In section~\ref{sec:BHNSResults}, we
create a number of initial data sets to demonstrate the robustness of
our solver by constructing BH-NS initial data sets with various values
of neutron star spin, black hole spin, and mass ratio. We conclude
with a discussion in section~\ref{sec:Conc}. Throughout this article
we use units where $G=c=M_{\odot}=1$.

\section{Initial Data Formalism}
\label{sec:BH-NSIDF}

In this section we will discuss the formalism used to solve the
Einstein field equations and create quasi-equilibrium initial data for
BH-NS binaries with spinning neutron stars. We employ the
extended-conformal thin-sandwich
formalism~\cite{Pfeiffer2003b,York1999} to cast the Einstein
constraint equations as a set of elliptic equations. Neutron star
spin is incorporated with the approach developed
in~\cite{Tichy:2011gw}, and the equations are solved by a
generalization of the initial data solver developed
in~\cite{FoucartEtAl:2008}.

We begin with the $3+1$ decomposition of the space-time metric
tensor,
\begin{equation}
g_{\mu\nu}dx^{\mu}dx^{\nu} = -\alpha^2dt^2 + \gamma_{ij}\left(dx^i +
  \beta^idt\right)\left(dx^j+\beta^jdt\right),
\end{equation}
where $\alpha$ is the lapse function, $\beta^i$ is the shift vector,
and $\gamma_{ij}$ is the induced metric on a spatial hypersurface
$\Sigma(t)$. The normal vector $n^{\mu}$ to $\Sigma(t)$ is related to
the coordinate time $t$ by $ t^{\mu} = \alpha n^{\mu} + \beta^{\mu}$.
The extrinsic curvature of $\Sigma(t)$ is given by
$ K_{\mu\nu} = -\frac{1}{2}\mathcal{L}_n\gamma_{\mu\nu}$, where
$\gamma_{\mu\nu}=g_{\mu\nu}+n_{\mu}n_{\nu}$ and $\mathcal{L}_n$ is the
Lie derivative in the direction of $n^{\mu}$. By construction
$K_{\mu\nu} $ is a purely spatial tensor by construction,
i.e. $K_{\mu\nu}n^\mu=0=K_{\nu\mu}n^\mu$, and so we restrict our
attention to the spatial part of the extrinsic curvature, $K^{ij}$. It
is convenient to decompose it into its trace and trace-free parts,
\begin{equation}
K^{ij} = A^{ij}+\frac{1}{3}K\gamma_{ij}.
\end{equation}
The matter in the system is modelled with the stress-energy tensor of
a perfect fluid 
\begin{equation}
T_{\mu\nu}=\left(\rho+P\right)u_{\mu}u_{\nu}+Pg_{\mu\nu},
\end{equation}
where $\rho$ is the fluid's energy density, $P$ is its pressure, and
$u^{\mu}$ is its four-velocity. It is further useful to define the
projections of the matter quantities,
\begin{eqnarray}
E &=& T^{\mu\nu}n_{\mu}n_{\nu},\\
S &=& \gamma^{ij}\gamma_{i\mu}\gamma_{j\nu}T^{\mu\nu}, \\
J^i &=& -\gamma^{i}_{\mu}T^{\mu\nu}n_{\nu}.
\end{eqnarray}
The spatial metric is conformally scaled,
\begin{equation}
\gamma_{ij}=\Psi^4\tilde{\gamma}_{ij},
\end{equation}
where $\Psi$ denotes the conformal factor and $\tilde{\gamma}_{ij}$
the conformal metric. Other quantities are conformally scaled as follows:
\begin{eqnarray}
E &=& \Psi^{-6}\tilde{E}, \\
S &=& \Psi^{-6}\tilde{S}, \\
J^i &=& \Psi^{-6}\tilde{J}^i, \\
A^{ij} &=& \Psi^{-10}\tilde{A}^{ij}, \\
\alpha &=& \Psi^{6}\tilde{\alpha}. 
\end{eqnarray}
$\tilde{A}^{ij}$ is related to the shift and to the time derivative of
the conformal metric, $\tilde{u}_{ij}=\partial_t\tilde{\gamma}_{ij}$,
by
\begin{equation}
\tilde{A}^{ij} =
\frac{1}{2\tilde{\alpha}}\left[\left(\tilde{\mathrm{L}}\beta\right)^{ij}-\tilde{u}^{ij}\right],
\end{equation}
where $\tilde{L}$ is the conformal longitudinal operator,
\begin{equation}
\left(\tilde{L}V\right)^{ij}=\tilde{\nabla}^iV^j + \tilde{\nabla}^jV^i
- \frac{2}{3}\tilde{\gamma}^{ij}\tilde{\nabla}_kV^k.
\end{equation}
With these definitions and conformal rescalings, the Einstein
constraint equations, and the Einstein evolution equation for the
trace of the extrinsic curvature yield a set of five coupled elliptic
equations, called the extended conformal thin sandwich (XCTS)
equations~\cite{Pfeiffer2003b}. They are written in the form

\begin{eqnarray}
\label{eq:XCTS-ConformalFactor}
\fl
\tilde{\nabla}^2\Psi - \frac{1}{8}\Psi\tilde{R} -
\frac{1}{12}\Psi^5K^2 
+\frac{1}{8}\Psi^{-7}\tilde{A}_{ij}\tilde{A}^{ij} &=-2\pi\Psi^{-1}\tilde{E},\\
\label{eq:XCTS-Shift}
\fl
2\tilde{\alpha}\bigg[\tilde{\nabla}_j\left(\frac{1}{2\tilde{\alpha}}\big(\tilde{L}\beta\big)^{ij}\right)-\tilde{\nabla}_j\left(\frac{1}{2\tilde{\alpha}}\tilde{u}^{ij}\right)
-\frac{2}{3}\Psi^6\tilde{\nabla}^iK\bigg] &=16\pi\tilde\alpha\Psi^4\tilde{J}^i,\\
\fl 
\tilde{\nabla}^2\left(\tilde{\alpha}\Psi^7\right) -
\left(\tilde{\alpha}\Psi^7\right)\bigg[\frac{1}{8}\tilde{R}+\frac{5}{12}\Psi^4K^2+\frac{7}{8}\Psi^{-8}\tilde{A}_{ij}\tilde{A}^{ij} 
\bigg] \nonumber \\
\qquad\qquad\qquad\qquad+\Psi^5\left(\partial_{t}K- \beta^{k}\partial_kK\right)&=-2\pi\tilde\alpha\Psi^{5}\big(\tilde{E}+2\tilde{S}\big).
\label{eq:XCTS-Lapse}
\end{eqnarray}
These equations are solved for the conformal factor, $\Psi$, the
shift, $\beta^i$, and the densitized lapse, $\tilde\alpha\Psi^7$.
Equations~(\ref{eq:XCTS-ConformalFactor})--(\ref{eq:XCTS-Lapse})
constitute the gravitational sector of the initial data construction.
The free data are $\tilde{\gamma}_{ij}$, $\tilde{u}_{ij}$, $K$ and
$\partial_t K$. Since we will be constructing initial data in a
corotating coordinate system, the free data corresponding to
time-derivatives can be naturally set to zero:
$\tilde{u}_{ij}=\partial_t K=0$. The choice of the conformal metric
and $K$ will be discussed in section~\ref{sec:BHNSNumMethods}.

Equations~(\ref{eq:XCTS-ConformalFactor})--(\ref{eq:XCTS-Lapse})
require boundary conditions at large separation, and at the excision
boundary of the black hole. At infinity\footnote{In practice we place
  the outer boundary of the computational grid at $R=10^{10}$.} are
the requirement of a Minkowski metric in the inertial frame
(\cite{FoucartEtAl:2008}):
\begin{eqnarray}
  \boldsymbol{\beta}_0&=&0,\label{eq:QQQQ}\\
  \alpha\Psi &=& 1,\\
  \Psi &=&1.
\end{eqnarray}
Here $\boldsymbol{\beta}_0$ is the shift in the inertial frame, which is related to the shift vector $\boldsymbol{\beta}$ by
\begin{equation}
\label{eq:ShiftExpansion}
\boldsymbol{\beta} = \boldsymbol{\beta}_0 + \boldsymbol{\Omega}\times{\bf r}+\dot{a}_0{\bf r},
\end{equation}
where $\boldsymbol{\Omega}$ is the orbital angular velocity of the system and $\dot{a}_0$ is a term used to give the system an infall velocity ${\bf v}=\dot{a}_0{\bf r}$.
The interior of the black hole is excised from the computation
domain. The boundary conditions at the surface of the black hole
apparent horizon, $\mathcal{H}$, are~\cite{Cook2004}:
\begin{eqnarray}
  \tilde{s}^k\nabla_k\log\Psi
  &=-\frac{1}{4}\left(\tilde{h}^{ij}\tilde{\nabla}_i\tilde{s}_j-\Psi^2h^{ij}K_{ij}\right)
\qquad &{\rm on} ~\mathcal{H}, \label{eq:BHBoundary}\\
\beta_{\perp}&=\alpha &{\rm on} ~\mathcal{H}, \label{eq:BHBoundary2} \\
\beta_{\parallel}^i&=\Omega_{j}^{BH}x_k\epsilon^{ijk}
&{\rm on}
~\mathcal{H}, \label{eq:BHBoundary3}.
\end{eqnarray}
In Eq.~(\ref{eq:BHBoundary}), $s^i=\Psi^{-2}\tilde{s}^i$ denotes the
outward pointing unit normal to the apparent horizon surface and
$h^{ij}=\gamma^{ih}-s^is^j$ is the induced metric on the surface. In
Eq.~(\ref{eq:BHBoundary3}), $\epsilon^{ijk}=\{\pm1,0\}$, is the totally
anti-symmetric symbol, $x_i$ are the Cartesian coordinates relative to
the center of the black hole and $\Omega_j^{BH}$ is a free vector that
determines the spin of the black hole.

Let us next focus on the matter content of the neutron star, which
enters through $\tilde{E}$, $\tilde{S}$, and $\tilde{J}^i$. The
energy density of the fluid is $\rho=\rho_0\left(1+\epsilon\right)$,
where $\rho_0$ is the baryon density and $\epsilon$ is the internal
energy. The specific enthalpy of the fluid is
\begin{equation}\label{eq:hDefn}
h=1+\epsilon+\frac{P}{\rho_0}.
\end{equation}
It is convenient to introduce a three-velocity $U^\mu$ satisfying
\begin{eqnarray}
  U^\mu n_\mu&=0, \label{eq:QQQQ2}\\
  u^{\mu} &=& \gamma_n\left(n^{\mu}+U^\mu\right).
\end{eqnarray}
These conditions imply
\begin{equation}
\gamma_n = \left(1 - \gamma_{ij}U^iU^j\right)^{-1/2}.
\end{equation}
Furthermore, we introduce
\begin{eqnarray}
  U^i_0 &=& \frac{\beta^i}{\alpha}, \\
\gamma_0 &=& \left(1 - \gamma_{ij}U^i_0U^j_0\right)^{-1/2}, \\
\gamma &=& \gamma_n\gamma_0\left(1-\gamma_{ij}U^iU^j_0\right).
\label{eq:QQQQ3}
\end{eqnarray}

Following \cite{Tichy:2011gw}, the three-velocity is written as the sum
of an irrotational part (the gradient of a potential $\phi$) and a
rotational part $W^i$,
\begin{equation}
U^i =
\frac{\Psi^{-4}\tilde{\gamma}^{ij}}{h\gamma_n}\left(\partial_j\phi+W_j\right).
\end{equation}
$W^i$ is a freely chosen, divergence-free vector field in this
formalism; we will discuss the choice of $W^i$ in section~\ref{sec:BHNSNumMethods}.

The matter fluid must satisfy the continuity equation and the Euler
equation. 
Under the assumptions made in \cite{Tichy:2011gw}, the continuity equation is a second order elliptic equation for the potential $\phi$:
\begin{equation}
\label{eq:Continuity1}
\frac{\rho_0}{h}\nabla^{\mu}\nabla_{\mu}\phi+\left(\nabla^{\mu}\phi\right)\nabla_{\mu}\frac{\rho_0}{h}=0.
\end{equation}
This can be re-written as
\begin{eqnarray}
\label{eq:Continuity2}
\fl 
\rho_0\,\bigg\{\!\!-\tilde{\gamma}^{ij}\partial_i\big(\partial_j\phi+W_j\big)  &+& \frac{h\beta^i\Psi^4}{\alpha}\partial_i\gamma_n + hK\gamma_n\Psi^4+\Big[\tilde{\gamma}^{ij}\tilde{\Gamma}^k_{ij}+\gamma^{ik}\partial_i\big(\ln \frac{h}{\alpha\Psi^2}\big)\Big] 
\big(\partial_k\phi+W_k\big) \bigg\} \nonumber\\
&=&\tilde{\gamma}^{ij}\big(\partial_i\phi+W_i\big)\partial_j\rho_0 - \frac{h\gamma_n\beta^i\Psi^4}{\alpha}\partial_i\rho_0.
\label{eq:Continuity}
\end{eqnarray}
Turning to the Euler equation, it can be solved for the specific
enthalpy $h$ as shown in~\cite{Tichy:2011gw}:
\begin{equation}\label{eq:hSoln}
h = \sqrt{L^2 -
  \left(\nabla_i\phi+W_i\right)\left(\nabla^i\phi+W^i\right)},
\end{equation}
where
\begin{equation}
L^2 =
\frac{b+\sqrt{b^2-4\alpha^4\left(\left(\nabla_i\phi+W_i\right)W^i\right)^2}}{2\alpha^2}
\end{equation}
and
\begin{equation}
b =
\left(\beta^i\nabla_i\phi+C\right)^2+2\alpha^2\left(\nabla_i\phi+W_i\right)W^i.
\end{equation}

The boundary condition on $\phi$ at the surface of the neutron star are deduced from the $\rho_0\rightarrow 0$ limit of the continuity equation:
\begin{equation}
\tilde{\gamma}^{ij}\left(\partial_i\phi+W_i\right)\partial_j\rho_0=\frac{h\gamma_n\beta^i\Psi^4}{\alpha}\partial_i\rho_0.
\end{equation}
Note that $\phi$ is only solved for inside the neutron stars, while
the metric variables are solved for everywhere.

The force balance equation at the center of the neutron star, $c^i$ is 
\begin{equation}
\nabla\log h=0 \qquad {\rm at}~x^i=c^i.
\end{equation}
We can re-write this equation as~\cite{Tichy:2011gw} 
 \begin{equation}
\label{eq:OmegaDriver}
\nabla\ln\left(\alpha^2-\gamma_{ij}\beta^{i}\beta^{j}\right)=-2\nabla\ln\Gamma,
\end{equation}
where
\begin{equation}
\Gamma
=\frac{\gamma_n\left(1-\left(\beta^i+\frac{W^i\alpha}{h\gamma_n}\right)\frac{\nabla_i\phi}{\alpha
    h\gamma_n}- \frac{W_i W^i}{\alpha^2\gamma_n^2}\right) } { \sqrt{ 1
    - \left(\frac{\beta^i}{\alpha}+\frac{W^i}{h\gamma_n}\right)
    \left(\frac{\beta_i}{\alpha}+\frac{W_i}{h\gamma_n}\right) } }.
\end{equation}
Since $\beta^i=\beta^i_0 +
\vec{\Omega}\times\vec{r} + \dot{a}\vec{r}$, where $\beta^i_0$ is the
shift in the inertial frame,
this is a second order equation for the orbital
frequency $\Omega$, when $\Gamma$ is held constant. If desired, this equation can be solved to find a
best guess for the orbital frequency. Alternatively, eccentricity
removal techniques, such as those used in \cite{Buonanno:2010yk,Tacik:2015tja} can be
used to find the best value of the orbital frequency.

$W^i$ is chosen as as to give the NS a uniform rotational
profile. Following our work in \cite{Tacik:2015tja}, we use
\begin{equation}
\label{eq:RotationTerm}
W^i=\epsilon^{ijk}\omega^jr^k,
\end{equation}
where $r^k$ is the position vector relative to the center of the star,
and $\omega^j$ is a freely chosen constant vector. Outside a radius
larger than the neutron star size, $W^i$ is set to zero, to avoid low
density material at high radius leading to spurious large velocities.
This is particularly important for large neutron star spins, or when
the black hole mass is much larger than the neutron star mass.

\section{Numerical Methods}
\label{sec:BHNSNumMethods}

\subsection{Domain decomposition}

The XCTS equations~\ref{eq:XCTS-ConformalFactor},~\ref{eq:XCTS-Shift},
and~\ref{eq:XCTS-Lapse}, combined with the continuity
equation~\ref{eq:Continuity2} form a set of six non-linear coupled
elliptic equations that must be solved. We use the pseudo-spectral
multi-domain elliptic solver developed in \cite{Pfeiffer2003} and
enhanced to incorporate matter
in~\cite{FoucartEtAl:2008,Tacik:2015tja,Haas:2016}.

\begin{figure}
\fboxsep0cm  
\centerline{\fbox{\includegraphics[width=0.8\textwidth]{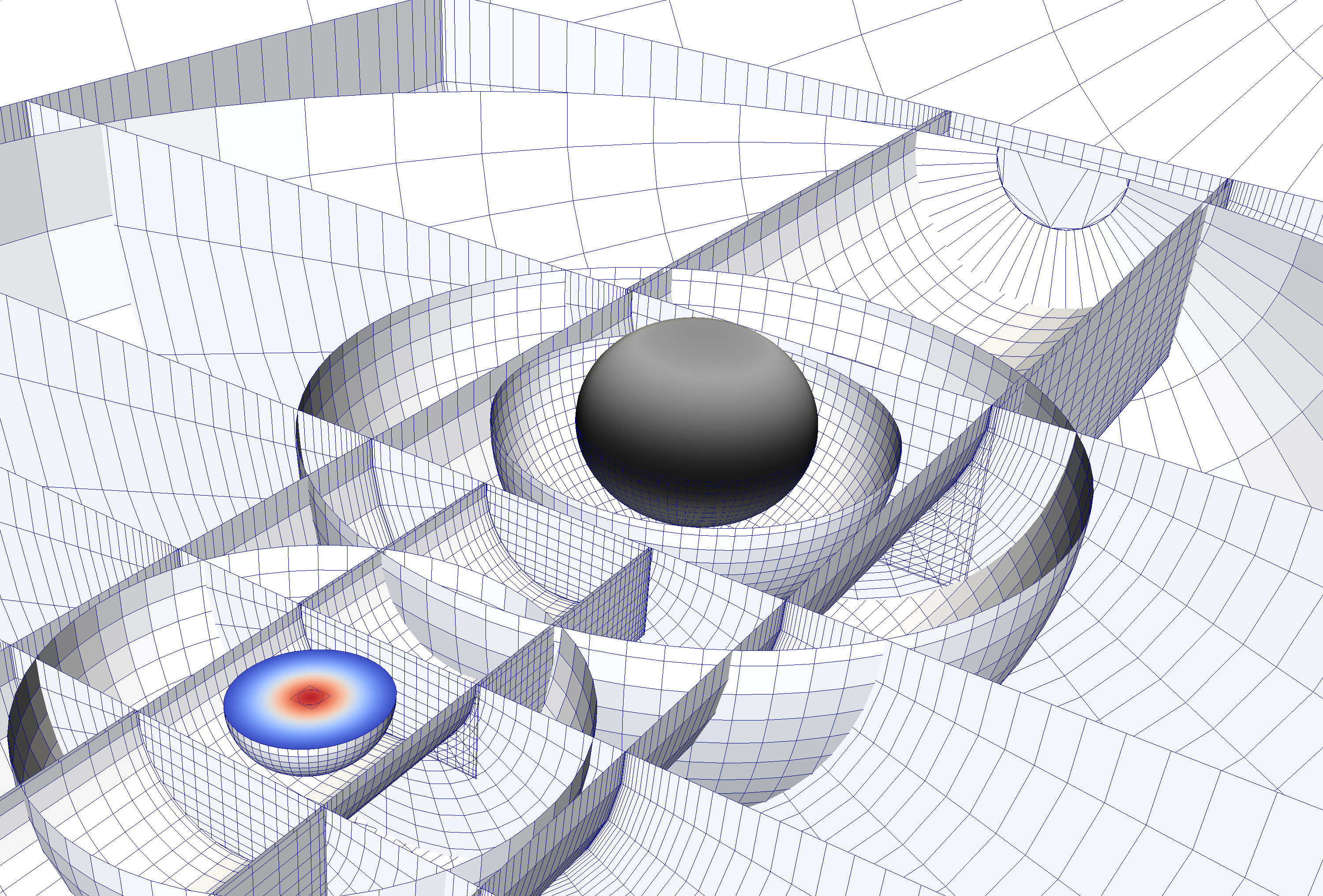}}}
\caption[Visualization of the BH-NS domain
decomposition.]{Visualization of the BH-NS domain decomposition. The object on the left is the neutron star, 
with the colours representing its density. The black object on the right represents the apparent horizon of the black hole. The blue wireframes
represent the various spheres, cylinders and rectangular parallelepipeds in the domain.

}
\label{fig:BHNSDomain}
\end{figure}

The computational domain has the black hole interior excised, and
extends to some large outer boundary ($R=10^{10}$ in practice). To
cover this computational domain with spectral expansions, we split the
domain into multiple subdomains as indicated in
figure~\ref{fig:BHNSDomain}, each one with its own spectral expansion:
The neutron star is covered by a spherical shell with outer boundary
deformed to coincide with the boundary of the neutron star
(cf. Eq.~(\ref{eq:NSSurf}) below). To avoid having to deal with
regularity conditions at the origin, this shell does not cover the
origin; rather a small cube is placed there which overlaps the
spherical shell. The neutron star is surrounded by one further
spherical shell. The black hole is surrounded by two concentric
spherical shells, where the inner boundary of the inner shell
coincides with the apparent horizon, where the boundary
conditions~(\ref{eq:BHBoundary})~--~(\ref{eq:BHBoundary3}) are imposed.
Three rectangular parallelepipeds surround the axis passing through the
centers of the BH and the NS - one between them and one on each side
of the objects. An additional eight cylindrical shells are placed
around the same axis to cover the intermediate field region. The
far-field region is covered by a large spherical shell whose outer
boundary is placed at $R=10^{10}$ using an inverse radial
mapping.

All variables (metric and hydrodynamical) are decomposed on sets of
basis functions on each subdomain. The type of basis function depends
on the topology on the subdomain. Finite difference schemes are needed
for hydrodynamical quantities during evolutions so as to capture
shocks, but for initial data, where shocks are not present, spectral
methods are suitable and exponential convergence can be achieved. The
resolution of each domain is synonymous with the number of colocation
points used. The resolution of each subdomain is initialized manually
at the start of the initial data solve; subsequently, the resolution
is adjusted several times using an adaptive mesh refinement (AMR) scheme (see
step~\ref{it:final}). To discuss the resolution of the computational
domain for the purpose of convergence tests, we denote by $N$
the total number of collocation points in all subdomains. $N^{1/3}$ is then a measure of linear resolution.
A typical initial data solve starts with, $N^{1/3}\sim 33$ and ends
with $N^{1/3}\sim 80$.

\subsection{Diagnostics}

The angular momentum of the black hole is
computed as~\cite{Lovelace2008,FoucartEtAl:2008}
\begin{equation}
\label{eq:BHSpin}
S=\frac{1}{8\pi}\oint_{\mathcal{H}}\phi^is^jK_{ij}dA.
\end{equation}
 In a space-time with aziumuthal symmetry, $\phi^i$ would represent
 the exact azimuthal Killing vector field generated by this symmetry.
Since azimuthal symmetry is not present in a binary system, we
instead use an {\it approximate} Killing vector. It is computed by solving a
shear minimization eigenvalue problem - see
\cite{Cook2007,Lovelace2008}  for details. The dimensionless spin is defined as
\begin{equation}
\label{eq:ChiDef}
\chi=\frac{S}{M^2},
\end{equation}
where $M$ is the Christodoulou mass,
\begin{equation}
M^2=M_{\rm irr}^2+\frac{S^2}{4M_{\rm irr}^2}.
\end{equation}
The irreducible mass is related to the surface area of the apparent horizon, $A$,
\begin{equation}\label{eq:Mirr}
M_{\rm irr}=\sqrt{A/16\pi}.
\end{equation}

We employ similar surface integrals to compute the dimensionless spin
of the neutron star. In particular we use Eq.~\ref{eq:BHSpin}, with
$\mathcal{H}$ replaced by the neutron star's surface, as defined in
Eq.~\ref{eq:NSSurf} to compute the star's angular momentum
$S_{\rm NS}$.

To define the neutron star's mass, we use the Arnowitt-Deser-Misner
(ADM) mass $M_{\rm ADM, NS}$ of an \emph{isolated} neutron star with
same rotation. In particular, we use the methods described
in~\cite{cook94a} to solve for the equilibrium state of an
isolated neutron star with the same baryon mass, equation of state,
and angular momentum, as the neutron star in our binary, and then
compute its ADM mass. The dimensionless neutron star spin is 
defined as
\begin{equation}
  \chi_{\rm NS}=\frac{S_{\rm NS}}{M_{\rm ADM, NS}^2}.
\end{equation}
Ref.~\cite{Tacik:2015tja} showed that this method of computing
neutron star spin was robust and accurate.

The ADM linear momentum is defined by a surface integral at infinity~\cite{ADM,York:1979},
\begin{equation}\label{eq:PADM}
P_{\rm ADM}^i = \frac{1}{8\pi}\oint_{S_{\infty}}K^{ij}dS_j.
\end{equation}
This integral relies on cancellation of leading order terms~\cite{FoucartEtAl:2008,Ossokine:2015yla}, which
results in loss of accuracy when evaluated at finite numerical
precision. Therefore, we use Gauss' law to rewrite~\cite{FoucartEtAl:2008,Ossokine:2015yla}
equation~(\ref{eq:PADM}) as a surface integral over a sphere with
smaller radius, $\mathcal{S}_0$, and a volume-integral over the volume
$V_0$ outside $\mathcal{S}_0$, 
\begin{equation}\label{eq:PADM2}
P_{\rm ADM}^i = \frac{1}{8\pi}\oint_{S_0}P^{ij}dS_j-\frac{1}{8\pi}\int_{V_0}G^idV,
\end{equation}
where
\begin{eqnarray}
P^{ij}=\Psi^{10}\left(K^{ij}-K\gamma^{ij}\right),\\
G^i=\tilde{\Gamma}^i_{jk}P^{jk}+\tilde{\Gamma}^j_{jk}P^{ik}-2\tilde{\gamma}_{jk}P^{jk}\tilde{\gamma}^{il}\partial_l\left(\log\Psi\right).
\end{eqnarray}

\subsection{Iterative procedure}

Construction of initial data begins by choosing the physical
parameters of the BH-NS binary, which we aim to achieve.
For the black hole, we specify:
\begin{itemize}
\item The black hole mass, $M_{\rm BH}$,
\item The black hole's dimensionless spin vector,
  $\vec{\chi}_{\rm BH}$.
\end{itemize}
For the neutron star we specify:
\begin{itemize}
\item The neutron star's baryon mass, $M_{b}$,
\item The neutron star's equation of state, 
\item The neutron star's spin vector, $\omega^i$.
\end{itemize}
Finally, characterizing the orbit are:
\begin{itemize}
\item The separation between the centers of the BH and NS, $D$,
\item The orbital angular velocity, $\Omega_0$,
\item The initial infall velocity parameter, $\dot{a}_0$.
\end{itemize}

Additionally, a prescription is required for the free metric
variables, $\tilde{\gamma}_{ij}$ and $K$. Near the black hole, we
would like these variables to approach the spatial metric
$\gamma_{ij}^{\rm KS}$ and mean curvature $K^{\rm KS}$ of a single
rotating black hole in Kerr-Schild coordinates. Away from the black hole (most notably in the vicinity of the neutron star), we desire conformal flatness and maximal slicing. Overall, therefore, we set 
\begin{eqnarray}
  \tilde\gamma_{ij}&=\delta_{ij} + e^{(-r/w)^4}\left(g_{ij}^{\rm KS} -
  \delta_{ij}\right),\\ K &= e^{(-r/w)^4} K^{\rm KS},
\end{eqnarray}
where $r$ is the distance to the center of the black hole and $w$ is the
roll-off distance.

Once all physical parameters are specified an iterative procedure is
used to solve the various elliptic equations and additional
conditions, We proceed as follows:
\begin{enumerate}
  \renewcommand{\theenumi}{\arabic{enumi}}
\item Initialize two counters for nested iterative loops, $k=0$ and $n=0$. Here, $k$ represents the AMR resolution iterations, and $n$ represents iterations at constant AMR resolution.
  \item
  \label{it:1}
  If $k=0$, 
  at the first iteration (i.e. step~\ref{it:secondlast} has not been
  reached yet), set $\omega^i=0$. Otherwise set $\omega^i$ to its
  desired value, cf. above. This has been found to improve overall
  convergence, especially for high neutron star spins.
\item 
\label{it:solve}
Solve the non-linear XCTS equations~\ref{eq:XCTS-Shift}--\ref{eq:XCTS-Lapse} for the metric variables
  $\beta^i,\Psi,\alpha\Psi$ assuming the matter source
  terms are fixed. For $n=0$ this defines the metric variables $X^{(0)}$ at the 0-th iteration, where $X$ indicates each of the metric variables.
If $n\ge 1$, update the metric variables using a relaxation
  scheme
\begin{equation}
\label{eq:Relaxation}
X^{(n+1)}=\lambda X^{*} + (1-\lambda)X^{(n)},
\end{equation}
where $X^{*}$ is the
result found by solving the XCTS equations. We use $\lambda=0.3$.

\item If both the NS and BH have either aligned spin or zero spin,
  impose equatorial symmetry. This speeds up convergence and
  decreases computational cost.

\item
  \label{it:toplevelparamsolve}

 If $k\ge 4$ go directly to
  step~\ref{it:omega}. (We generally find that after four
  resolution-updates, the stellar and black hole parameters are
  computed to sufficient accuracy. Skipping
  steps~\ref{it:surface}--\ref{it:OmegaUpdate} decreases computational
  cost.)
\item 
\label{it:surface}
Locate the surface of the star. The surface of the star is
  represented in terms of spherical harmonics
\begin{equation}
\label{eq:NSSurf}
R(\theta,\phi)=\sum_{l=0}^{l_{\rm max}}\sum_{|m|\le l} c_{lm}Y^{lm}(\theta,\phi).
\end{equation}
The coefficients $c_{lm}$ are determined by solving the relation $h\left(R\left(\theta,\phi\right)\right)=1$.
We generally use $l_{\rm max}=11$.

\item Compute the ADM linear momentum $P_{\rm ADM}$ by evaluating Eq.~\ref{eq:PADM}.
If its norm has changed by less than 10\% in the last iteration, move
the center of the BH by an amount $\delta\vec{c}$, designed to zero the in-plane components of $P^i_{\rm ADM}$, by finding $\delta \vec{c}$ such that $\delta\vec{c}
\times \vec{\Omega}_0=\vec{P}_{\rm ADM}$. Additionally, increase the
radius of the excision surface $r_{\rm ex}$ to
drive $M_{\rm BH}$ to the desired value by applying (\cite{Buchman:2012dw})
\begin{equation}
\delta r_{\rm ex} = -r_{\rm ex} \frac{M_{\rm BH} - M_{\rm
    BH}^{*}}{M_{\rm BH}},
\end{equation}
where $M_{\rm BH}$ is the measured value in the initial data solve and
$M_{\rm BH}^{*}$ is the desired value.

\item \label{it:OmegaUpdate} Compute the spin of the BH by evaluating Eq.~\ref{eq:BHSpin}. Then
  modify the vector $\Omega^i_{\rm BH}$ in Eq.~\ref{eq:BHBoundary3} to drive
  the black hole spin to the target value,  by applying (\cite{Buchman:2012dw})
\begin{equation}
\delta \Omega_{\rm BH}^i = -\frac{\chi^i_{\rm BH}-\chi_{\rm
    BH}^{*i}}{4M}+\frac{M_{\rm BH}-M_{\rm BH}^{*}}{4M_{\rm
    BH}^2}\chi_{\rm BH}^i,
\end{equation}
where $\chi^i_{\rm BH}$ is the computed black hole spin, and $\chi_{\rm
    BH}^{*i}$ is the target spin (see also~\cite{Ossokine:2015yla}).

\item
\label{it:omega}
 If desired, adjust the orbital angular frequency using
  Eq.~\ref{eq:OmegaDriver}, which, after expanding the shift as 
$\beta^i=\beta^i_0 +
\vec{\Omega}\times\vec{r} + \dot{a}\vec{r}$, is a second-order
equation for $\Omega$.

\item Fix the Euler constant by evaluating the integral
\begin{equation}
M_{B}=\int \rho_0\Psi^6\gamma_ndV
\end{equation}
as a function of the Euler
constant $C$ (recall that $C$ enters into $h$, cf. Eq.~\ref{eq:hSoln}, and that $\rho_0$, in turn, depends on $h$, cf. Eq.~\ref{eq:hDefn}). With the secant method, find the value of $C$ that yields the desired baryon mass of the neutron star.

\item Solve the elliptic equation~(\ref{eq:Continuity2}) for the velocity potential $\phi$
  and update $\phi$ using the relaxation scheme in Eq.~\ref{eq:Relaxation}.
  
\item
\label{it:secondlast}
Check whether the Euler constant, black hole mass, black hole spin, ADM linear momentum, and the constraints 
are satisfied to the desired accuracy. If so, proceed to step~\ref{it:final}. Otherwise increment $n$ and return to step~\ref{it:solve}.

\item 
\label{it:final}
Compute the truncation error for the current solution by examining the
spectral coefficients of the metric variables~\cite{Szilagyi:2014fna}.
If the truncation error is too large (generally we use $10^{-9}$ as
the criterion), adjust the number of grid points. Then increment $k$, set $n=0$, and return to
step~\ref{it:1}. This adaptive refinement is based on the target
truncation error and the measured convergence rate of the
solution. See \cite{Szilagyi:2014fna} for a complete description of
this procedure.

\end{enumerate}

\section{Results}
\label{sec:BHNSResults}

\subsection{Initial Data Set Parameters}

Our primary goal is to establish the performance of the initial data
solver described in Sec.~\ref{sec:BHNSNumMethods}. The parameter
space of BH-NS binaries is large, encompassing the masses of black
hole and neutron star, their spin magnitudes and their spin
directions, as well as the compactness of the neutron star. As our
first stage in exploring this parameter space, we add neutron star
spin to BH-NS initial data sets that already appeared in the
literature before (namely in Foucart et al.~\cite{Foucart:2013a}). Our basis is six BH-NS configurations with
different compactnesses  of the neutron star, and with different
orientations of the black hole spin relative to
the orbital angular momentum. All base-configurations have mass ratio
$q=7$,
black hole spin magnitude of $\chi_{\rm BH}=0.9$ and
neutron star mass of
$M^{\rm NS}_{\rm ADM}=1.4M_{\odot}$ (recall that $M^{\rm NS}_{\rm ADM}$ is the mass of an individual neutron star of the same properties). 

We explore three polytropic equations of state, $P=\kappa\rho^\Gamma$, all with
$\Gamma=2$. $\kappa$ is chosen to achieve neutron star compactnesses
$C=R/M=0.170, 0.156$, $0.144$, and radii of approximately
$12{\rm km}$, $13{\rm km}$, $14{\rm km}$, respectively, for
non-spinning neutron stars with ADM Mass $1.4M_{\odot}$. 

For all three equations of state, we consider BH spin-direction
parallel to the orbital angular momentum. For the stiffest equation
of state, we also vary the BH-spin direction and compute initial data
sets for misalignment angles $\iota=20^{\circ}, 40^{\circ},$ and
$60^{\circ}$. The base-configurations are named {\tt R}xx{\tt i}yy,
where 'xx' denotes the approximate NS radius in kilometers and 'yy' denotes
the inclination between BH spin direction and the orbital angular
momentum in degrees (for instance {\tt R14i20}).

\begin{table}
\caption[Initial data set parameters for series of 36 BH-NS initial
data sets.]{\label{tab:36Sets} Full set of parameters of
 the 36 sets of initial data constructed here. 

Given are 
 angle between the black hole spin and the orbital angular
  momentum $\theta_{\rm BH}$, baryon mass of the neutron star  $M^B_{\rm NS}$, orbital frequency $M\Omega_0$, spin vector $\vec{\omega}_{\rm NS}$ of the neutron star (cf. Eq.~\ref{eq:RotationTerm}), and the dimensionless spin of the neutron star, $\vec{\chi}_{\rm NS}$.
}
\begin{indented}

\item[] \begin{tabular}{l c c S S[table-format=3.7,table-align-text-post=false] S[table-format=3.5,table-align-text-post=false]}
\br
Name & $\Theta_{\rm BH}$ & $M^B_{\rm NS}$ & $M\Omega_{0}$ & {$\vec{\omega}_{\rm NS}$} & $\chi_{\rm NS}$ 
\\
\mr
{\tt R12i0$\uparrow$}&$0^\circ$ & 1.5212 & 0.0413  & 0.00667{$\hat{z}$}\hfill & 0.0995 \\
{\tt R12i0$\Uparrow$}&$0^\circ$ & 1.5212 & 0.0413  & 0.0225$\hat{z}$ & 0.4093 \\
{\tt R12i0$\downarrow$}&$0^\circ$ & 1.5212 & 0.0413  & -0.00667$\hat{z}$& -0.0895\\
{\tt R12i0$\Downarrow$}&$0^\circ$ & 1.5212 & 0.0413  & -0.0225$\hat{z}$ & -0.4030 \\
{\tt R12i0$\rightarrow$}&$0^\circ$ & 1.5212 & 0.0413  & 0.00667$\hat{x}$ & 0.0936\\
{\tt R12i0$\Rightarrow$}&$0^\circ$ & 1.5212 & 0.0413  & 0.0225$\hat{x}$ & 0.3989 \\
\mr
{\tt R13i0$\uparrow$}&$0^\circ$ & 1.5128 & 0.0413   & 0.00555$\hat{z}$ & 0.0997 \\
{\tt R13i0$\Uparrow$}&$0^\circ$ & 1.5128 & 0.0413  & 0.019$\hat{z}$ & 0.3911 \\
{\tt R13i0$\downarrow$}&$0^\circ$ & 1.5128 & 0.0413  & -0.00555$\hat{z}$ & -0.0845\\
{\tt R13i0$\Downarrow$}&$0^\circ$ & 1.5128 & 0.0413  & -0.019$\hat{z}$ & -0.3793 \\
{\tt R13i0$\rightarrow$}&$0^\circ$ & 1.5128 & 0.0413 &  0.00555$\hat{x}$ & 0.0913\\
{\tt R13i0$\Rightarrow$}&$0^\circ$ & 1.5128 & 0.0413  & 0.019$\hat{x}$ & 0.3771 \\
\mr
{\tt R14i0$\uparrow$}&$0^\circ$ & 1.5049 & 0.0413 &  0.005541$\hat{z}$ & 0.1188\\
{\tt R14i0$\Uparrow$}&$0^\circ$ & 1.5049 & 0.0413 &  0.017$\hat{z}$ & 0.4109\\
{\tt R14i0$\downarrow$}&$0^\circ$ & 1.5049 & 0.0413 & -0.005541$\hat{z}$& -0.0965\\
{\tt R14i0$\Downarrow$}&$0^\circ$ & 1.5049 & 0.0413  & -0.017$\hat{z}$ & -0.3915\\
{\tt R14i0$\rightarrow$}&$0^\circ$ & 1.5049 & 0.0413  & 0.005541$\hat{x}$ & 0.1066\\
{\tt R14i0$\Rightarrow$}&$0^\circ$ & 1.5049 & 0.0413  & 0.017$\hat{x}$ & 0.3907\\
\mr
{\tt R14i20$\uparrow$}&$20^\circ$ & 1.5049 & 0.0412  & 0.005541$\hat{z}$ & 0.1188 \\
{\tt R14i20$\Uparrow$}&$20^\circ$ & 1.5049 & 0.0412  & 0.017$\hat{z}$ & 0.4110\\
{\tt R14i20$\downarrow$}&$20^\circ$ & 1.5049 & 0.0412  &  -0.005541$\hat{z}$& -0.0964\\
{\tt R14i20$\Downarrow$}&$20^\circ$ & 1.5049 & 0.0412  & -0.017$\hat{z}$ & -0.3915\\
{\tt R14i20$\rightarrow$}&$20^\circ$ & 1.5049 & 0.0412  & 0.005541$\hat{x}$ & 0.1064\\
{\tt R14i20$\Rightarrow$}&$20^\circ$ & 1.5049 & 0.0412  & 0.017$\hat{x}$ & 0.3905 \\
\mr
{\tt R14i40$\uparrow$}&$40^\circ$ & 1.5049 & 0.0412  &  0.005541$\hat{z}$ & 0.1193 \\
{\tt R14i40$\Uparrow$}&$40^\circ$ & 1.5049 & 0.0412  & 0.017$\hat{z}$ & 0.4117\\
{\tt R14i40$\downarrow$}&$40^\circ$ & 1.5049 & 0.0412  & -0.005541$\hat{z}$& -0.0961\\
{\tt R14i40$\Downarrow$}&$40^\circ$ & 1.5049 & 0.0412  & -0.017$\hat{z}$ & -0.3908\\
{\tt R14i40$\rightarrow$}&$40^\circ$ & 1.5049 & 0.0412 & 0.005541$\hat{x}$ & 0.1064\\
{\tt R14i40$\Rightarrow$}&$40^\circ$ & 1.5049 & 0.0412  & 0.017$\hat{x}$ & 0.3905 \\
\mr
{\tt R14i60$\uparrow$}&$60^\circ$ & 1.5049 & 0.0415  &  0.005541$\hat{z}$ & 0.1200 \\
{\tt R14i60$\Uparrow$}&$60^\circ$ & 1.5049 & 0.0415  & 0.017$\hat{z}$ & 0.4132\\
{\tt R14i60$\downarrow$}&$60^\circ$ & 1.5049 & 0.0415  & -0.005541$\hat{z}$& -0.0954\\
{\tt R14i60$\Downarrow$}&$60^\circ$ & 1.5049 & 0.0415 & -0.017$\hat{z}$ & -0.3898\\
{\tt R14i60$\rightarrow$}&$60^\circ$ & 1.5049 & 0.0415 & 0.005541$\hat{x}$ & 0.1061\\
{\tt R14i60$\Rightarrow$}&$60^\circ$ & 1.5049 & 0.0415 & 0.017$\hat{x}$ & 0.3903\\
\br
\end{tabular}
\end{indented}
\end{table}

For the base-configurations, the following secondary choices are made:
The non-parallel part of the black hole spin is set parallel
to the $\hat{x}$ axis, i.e., the approximate axis between the BH and the NS. In each case the initial separation between the
black hole and the neutron star is $D=7.44M$, where
$M=M_{\rm BH}+M^{\rm ADM}_{\rm NS}$ is the total mass of the
binary. The initial infall velocity parameter $\dot{a}_0$ is set to
$0$. The orbital angular velocity, $\Omega_0$, is the same as in
\cite{Foucart:2013a} and is indicated in
table~\ref{tab:36Sets}. The above constitutes 6 different
configurations.

We combine each of the six base-configurations with six different
configurations of neutron star spins for a total of 36 total
configurations. In particular we choose three directions - aligned
with the orbital angular momentum, anti-aligned with the orbital
angular momentum, and parallel to the orbital plane (along the
$+\hat{x}$ direction). For each of these three $\hat\chi_{\rm NS}$ directions, we consider
``large'' and ``small'' neutron star spin magnitude,
$\chi_{\rm NS}\sim 0.4$ and $\chi_{\rm NS}\sim 0.1$. In our
naming notation, we use a double arrow ($\Uparrow$) for the
large $\chi_{\rm NS}$ configurations and a single arrow ($\uparrow$) for the
small $\chi_{\rm NS}$ configurations, with the direction of the arrow indicating the direction
of the NS spin vector (e.g. {\tt R14i20}$\rightarrow$). The full parameters of the initial data sets
are summarized in Table ~\ref{tab:36Sets}.

In the present paper, we do not perform evolutions of these initial
data sets. Based on the evolutions in~\cite{Foucart:2013a} (for
non-spinning NS), we expect these initial data sets to correspond to
binaries that will proceed through $\sim 7$ to $\sim 10$ orbits before
merger.

\subsection{Convergence of the Initial Data Solver}
\label{sec:Convergence}

To assess the convergence of the initial data solver we will begin by looking at the convergence of the iterative part of the solver. That is, the convergence of steps 1-12 in the iterative procedure
described above. We will first focus on one particular initial data
set of the 36 in Tab~\ref{tab:36Sets} - namely the {\tt
  R14i60$\Uparrow$} initial data set. The results we present for {\tt
  R14i60$\Uparrow$}, however, are representative for all of the 36 sets considered.

\begin{figure}
\centerline{\includegraphics[scale=0.7]{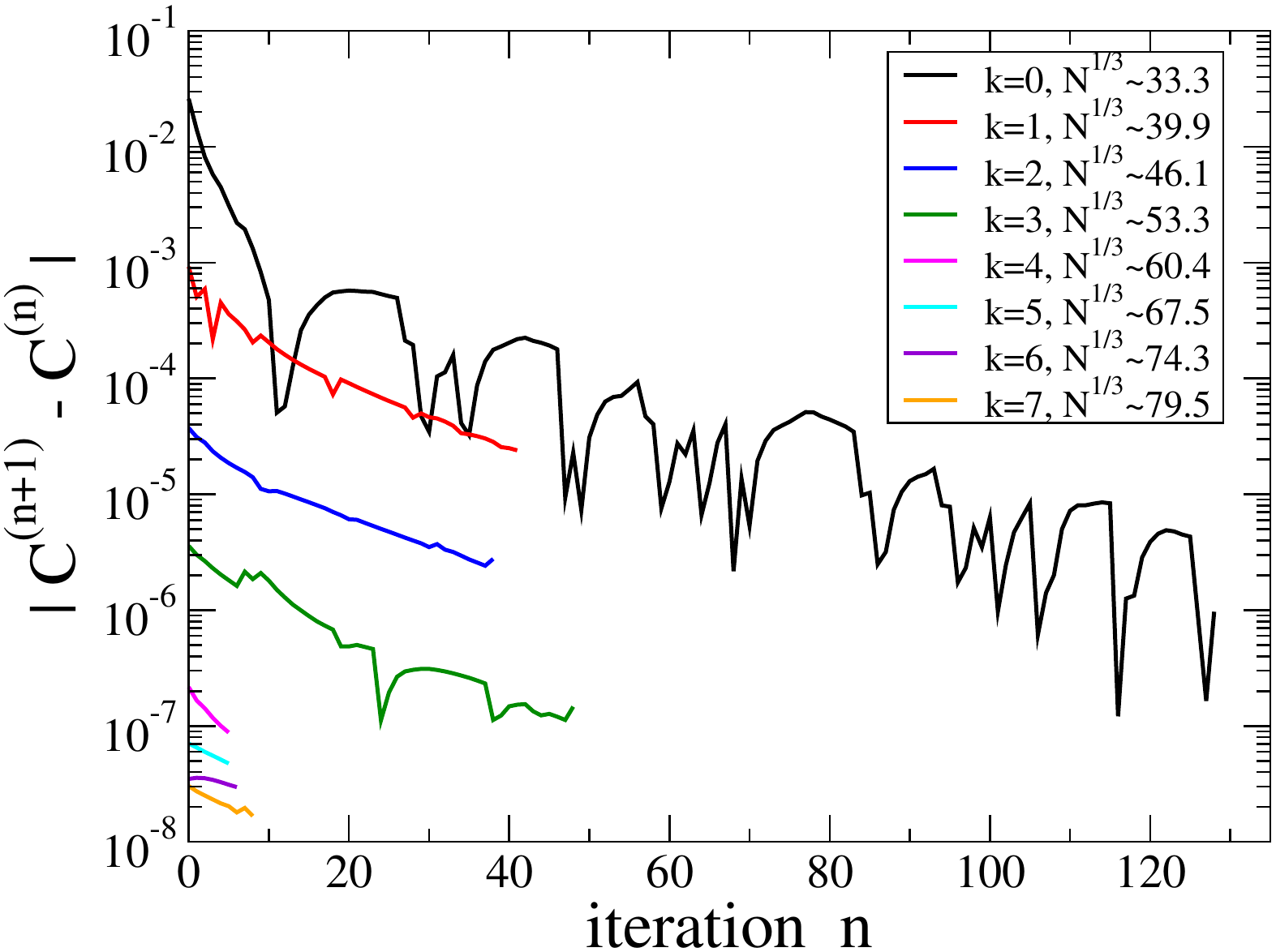}}
\caption{\label{Fig:EulerConv}
Absolute difference between neighbouring iterations of the Euler
constant for the {\tt R14i60$\Uparrow$} initial data set. $k$ labels AMR adjustment iterations, and $n$ the inner iterative loop at fixed grid-resolution. }
\end{figure}

We begin by looking at the convergence of the Euler constant, $C$. In figure~\ref{Fig:EulerConv} we plot the absolute difference in $C$ between neighbouring iterations for the eight
different resolutions used in the initial data solve. 
In the figure we see that at a given resolution these differences
decrease exponentially with iteration as expected for the relaxation
scheme employed (cf. Eq.~\ref{eq:Relaxation}). Meanwhile the differences also decrease with increasing resolution. 
We find similar results for all the other initial data sets we
consider.

\begin{figure}
  \centerline{\includegraphics[scale=0.7]{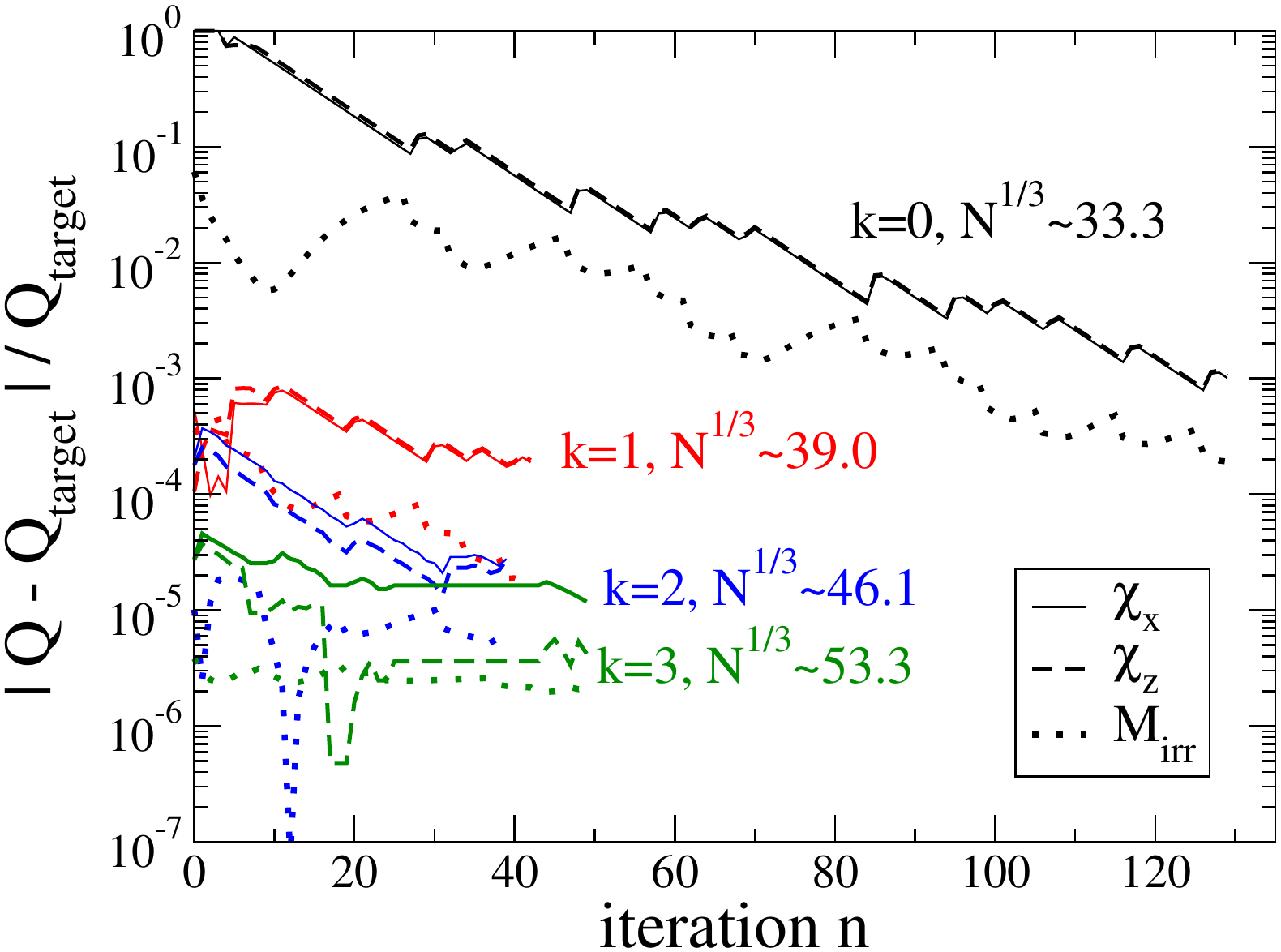}}
\caption[Convergence of black hole spin and mass.]
{\label{Fig:BHSpinConv}Fractional difference from desired values of
  black hole spin and mass. Shown is the solution of {\tt
    R14i60$\Uparrow$} initial data set as a function of iteration
  count. The four colours represent the four different resolutions
  $k=0,\ldots,3$ at which the black hole spin is measured. }
\end{figure}

Next, we will look at the properties of the black hole to verify that
they converge as expected in the presence of a spinning neutron star,
using again {\tt R14i60$\Uparrow$} as our example. We focus on the
black hole spin $\vec\chi_{\rm BH}$ which is controlled by the
parameter $\Omega_j^{\rm BH}$ in Eq.~\ref{eq:BHBoundary3}, and the
irreducible black hole mass, $M_{\rm irr}$ (cf. Eq.~\ref{eq:Mirr}).

Figure~\ref{Fig:BHSpinConv} shows the fractional difference for these quantities to their desired target value.
 The
difference is plotted as a funciton of iteration, for four different
resolutions. Recall that the BH spin is only adjusted on iterations $k=0,1,2,3$, and not thereafter (cf. Step~\ref{it:toplevelparamsolve}). In general we see a decrease in this difference with iteration, especially at the first resolution, therefore showing that the iterative solver is correctly driving the the black hole properties to the target values.
Furthermore, that this difference decreases with resolution, and
we are able to achieve an accuracy of about $10^{-5}$ in the BH spin
and mass. Note that these differences continue to remain small for $k>3$.

\begin{figure}
\centerline{\includegraphics[scale=0.7]{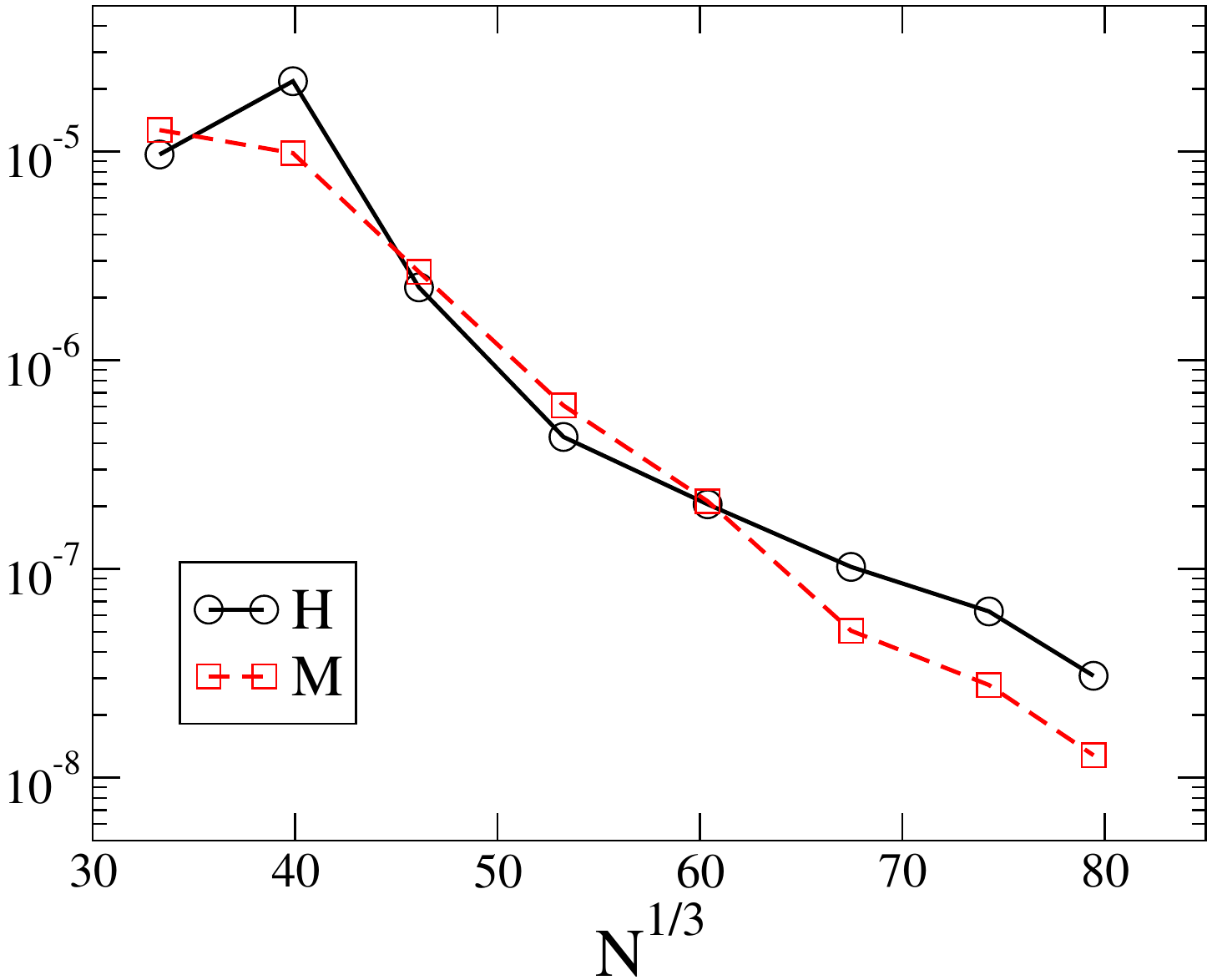}
}
\caption[Hamiltonian and momentum constraints of the {\tt R14i60$\Uparrow$} ID set]{\label{fig:HamMom} The Hamiltonian and momentum constraints for the {\tt R14i60$\Uparrow$} initial data set
as a function of resolution. We find exponential convergence in both.}
\end{figure}

Having established the convergence of the iterative procedure, we
turn now to the global
properties of the solution, continuing to focus on the {\tt R14i60$\Uparrow$} ID set. We first consider the Hamiltonian and momentum constraints, computed as
\begin{eqnarray}
H&=||\frac{R_{\Psi}}{8\Psi^5}||,\\
M &= ||\frac{R_{\beta}}{2\alpha\Psi^4}||,
\end{eqnarray}
where $R_{\Psi}$ and $R_{\beta}$ are the residuals of
Eqs.~\ref{eq:XCTS-ConformalFactor} and~\ref{eq:XCTS-Shift},
respectively, and $||.||$ represents the $L2$ norm over all
collocation points of the computational domain. The constraints for
this ID set are shown in figure~\ref{fig:HamMom}. We find exponential
convergence in the constraints, as expected for spectral methods. 
The increase at the second iteration ($k=1$) arises because the
neutron star spin is only activated in the second iteration 
(cf. step~\ref{it:1}).

\begin{figure}
\centerline{\includegraphics[scale=0.7]{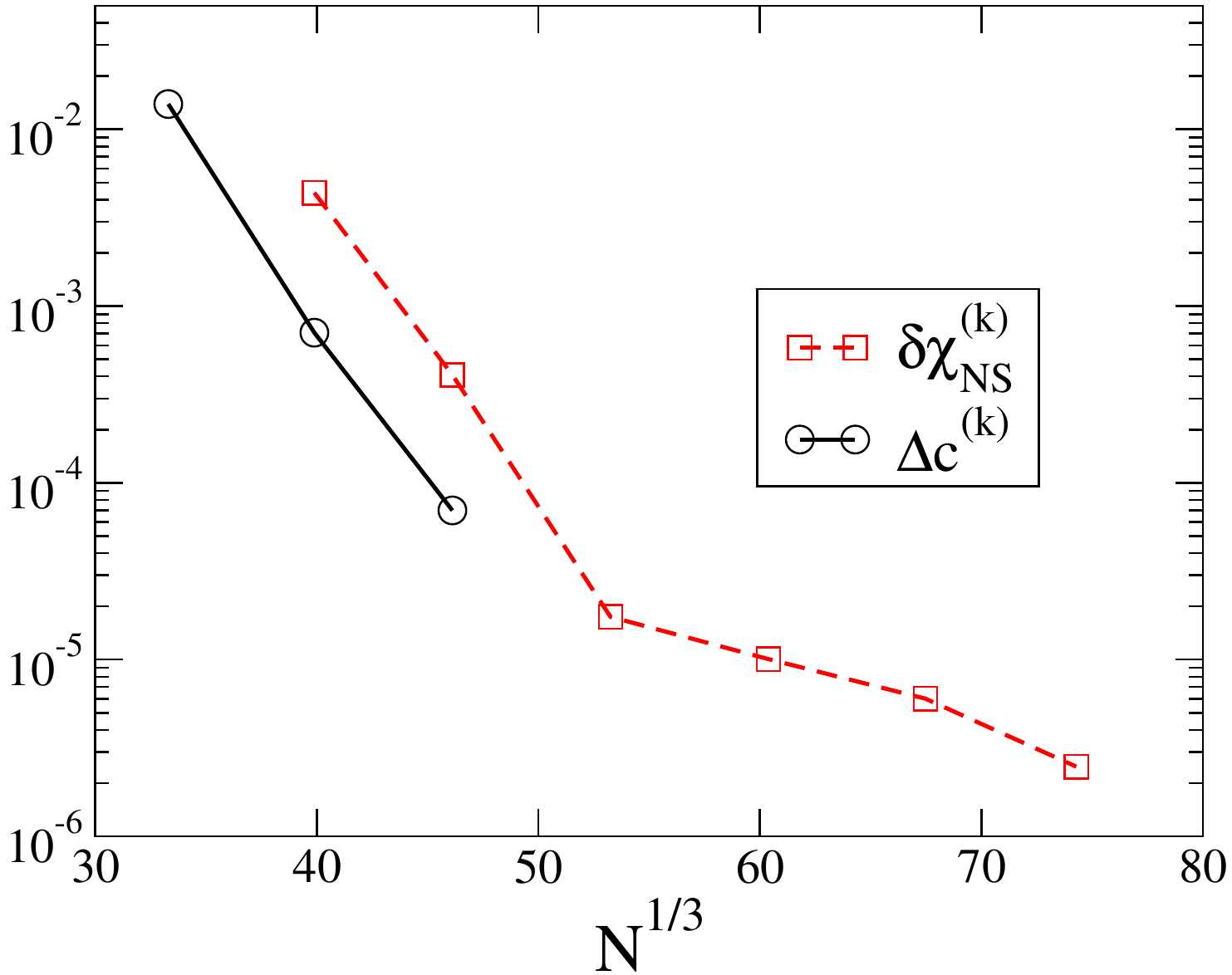}}
\caption[Neutron star surface and spin accuracy.]{\label{fig:SpinDiff}
  Accuracy of neutron star properties for the solution of {\tt R14i60$\Uparrow$}. Plotted are the accuracy of the
  NS surface, $\Delta c^{(k)}$, as defined in Eq.~\ref{eq:NSSurf2} and
  the fractional accuracy of the NS spin (equation~\ref{eq:NSspin2}).
}
\end{figure}

Finally, we look at the properties of the neutron star. As noted in Eq.~\ref{eq:NSSurf},
the neutron star surface is expressed as a sum of spherical
harmonics.
To evaluate the convergence of the surface location, we define the
quantity
\begin{equation}
\label{eq:NSSurf2}
\Delta c^{(k)}=\sqrt{\sum_{l,m}
\left(c_{lm}^{(k)}-c^{(k_{\rm max})}_{lm}\right)^2},
\end{equation}
where $k$ represents the current resolution, and $k_{\rm max}$ represents the
highest resolution. This quantity is plotted in
figure~\ref{fig:SpinDiff}. Similar to the black hole surface, the
neutron star surface is only computed for the first four resolutions,
and so we have three data points shown. We find exponential
convergence in this quantity. We also look at the convergence of the
neutron star spin $\chi_{\rm NS}$ measured at each resolution. In
figure~\ref{fig:SpinDiff}, we plot the fractional difference in
$\chi_{\rm NS}$ between neighbouring resolutions. That is, we plot
\begin{equation}\label{eq:NSspin2}
\delta\chi^{(k)}_{\rm NS}=\frac{\big| \chi^{(k+1)}_{\rm NS}-\chi^{(k)}_{\rm NS} \big|}{\chi^{(k)}_{\rm NS}}.
\end{equation}
Figure~\ref{fig:SpinDiff} exhibits exponential convergence, although
there are two disinctly different slopes in the data, once we cease to
update the NS surface for $k\ge 4$. Nevertheless, we are able to
measure the spin to an accuracy of about $10^{-6}$. We have omitted
the first data point of $\delta\chi^{(k)}_{\rm NS}$, because the 
NS spin is not activated for $k=0$ (cf. Step.~ \ref{it:1}).

The above data all show that we have established the convergence of
our initial data solver, by showing exponential convergence of the
iterative solver, the black hole properties, neutron star properties,
and the constraints.

\subsection{Broader exploration of parameter space}
\label{sec:MoreParameters}

All initial-data sets constructed so far share the same black hole
mass and black hole spin-magnitude, $M_{\rm BH}=9.8_\odot$
and $\chi_{\rm BH}=0.9$. In this section, we relax these
restrictions, and also explore the range of possible neutron star
spins our code is capable to construct. In total, we consider three additional sequences of initial-data sets:

\begin{figure}
  \centerline{\includegraphics[width=0.7\columnwidth,clip=true, trim=90
    47 60 68]{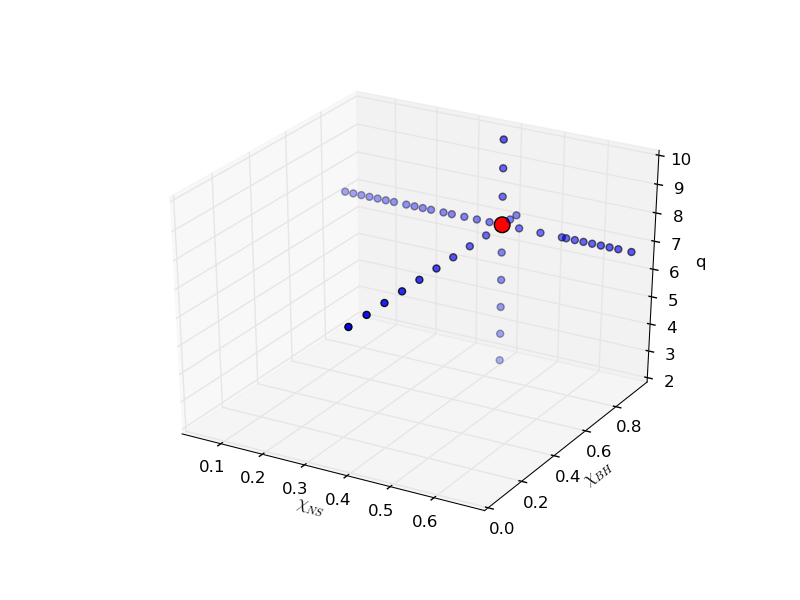} }
\caption[3d parameter space plot of BH-NS initial data
  sets.]{\label{fig:3dparam} Parameter space exploration. Starting
  from {\tt R14i0$\Uparrow$}(large red circle) we vary (i) the BH
  spin $\chi_{\rm BH}$, (ii) the NS spin $\chi_{\rm NS}$ and (iii) the
  black hole mass $M_{\rm BH}$, indicated by the mass-ratio $q$.}
\end{figure}

\begin{figure}
\centerline{\includegraphics[scale=0.7]{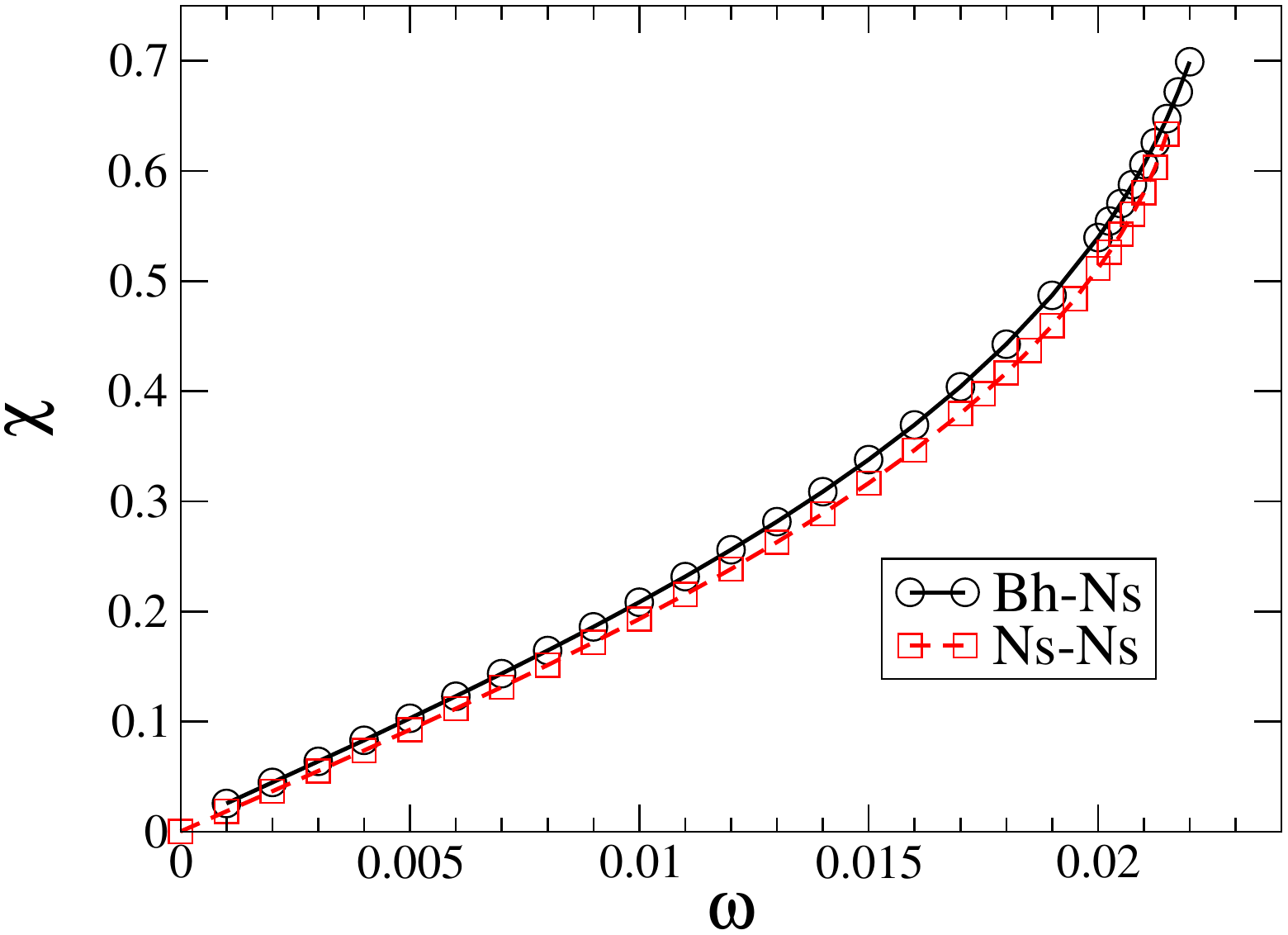}}
\caption[$\chi_{\rm NS}$ as a function of $\omega_{\rm NS}$ for bh-ns
binaries]
{\label{fig:ChiVOmega}
Neutron star spin $\chi$ as a function of neutron star spin parameter
$\omega$ for a sequence of initial data sets. The black hole spin is
constant at $\chi=0.9$ and the mass ratio is $q=7$. The dashed red curve is
from NS-NS binaries, with somewhat different neutron star parameters.}
\end{figure}

First, we consider a
sequence that varies the neutron star spin from $\chi_{\rm NS}=0$ to
$\chi_{\rm NS}\sim0.7$, keeping it aligned with the orbital angular
momentum. In these initial data sets, the other binary parameters are
the same as in the {\tt R14i0} runs. Namely, the neutron star mass,
equation of state, black hole mass, black hole spin, initial
separation and orbital angular frequency. Second, we consider a
sequence of runs where we vary the black hole spin from $\chi_{\rm
  BH}=0$ to $\chi_{\rm BH}=0.99$, while keeping the other binary
parameters as in the {\tt R14i0$\Uparrow$} run. Finally, we consider
a sequence of runs where we vary the mass ratio from $q=2$ to $q=10$.
Figure~\ref{fig:3dparam} summarizes all the initial data sets along
the axes of $\chi_{\rm NS}$, $\chi_{\rm BH}$ and $q$.

\begin{figure}
\centerline{\includegraphics[scale=0.7]{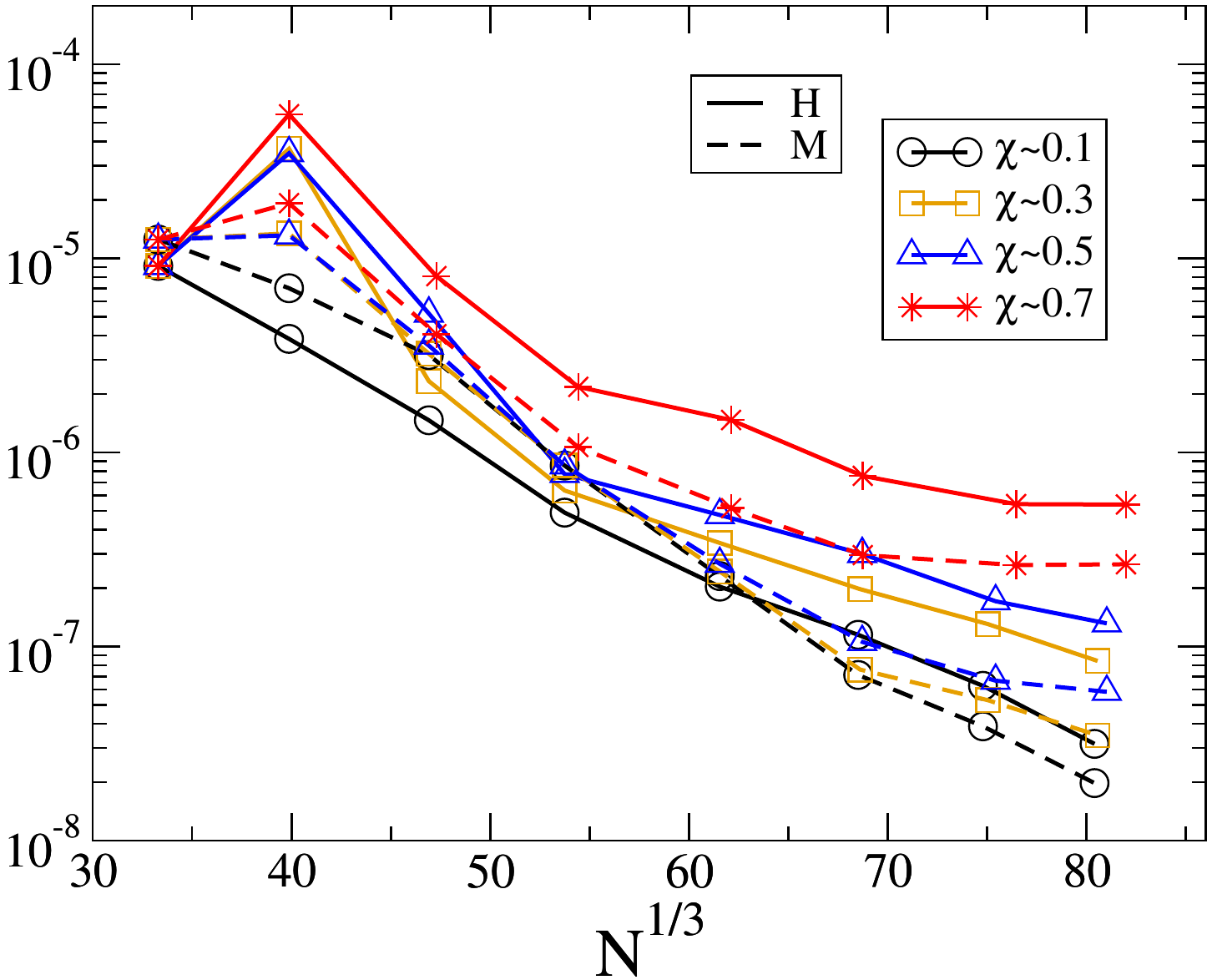}}
\caption[Constraints for the $\chi_{\rm NS}$ sequence.]{\label{fig:OmegaSeqHamMom} Hamiltonian (solid curves) and
momentum (dotted curves) constraints for four different neutron star spins.}
\end{figure}

We begin by varying the neutron star rotation parameter $\omega$. The
parameters in this sequence are otherwise the same as in the {\tt
  R14i0} data sets, and the neutron star spin is kept aligned with the
orbital angular momentum. In Fig.~\ref{fig:ChiVOmega} we plot the
measured neutron star spin $\chi_{\rm NS}$ as a function of the code
parameter $\omega$ for the full sequence. As expected, we find a
linear relationship at low $\omega$, but the relationship becomes
non-linear at higher $\omega$, as the neutron star's size, and thus
moment of inertia, becomes an appreciable function of spin. We find
that the solver breaks down around $\chi_{\rm NS}\sim 0.7$, which is the
maximum spin parameter for neutron stars found in~\cite{Lo:2010bj}. Figure~\ref{fig:ChiVOmega} also shows the
corresponding $\chi_{\rm NS}$ vs. $\omega$ curve for a binary neutron
star of mass-ratio $q=1$ with both stars carrying the same aligned
spin magnitude as presented in~\cite{Tacik:2015tja}. The NS-NS data
use different NS parameters, with mass $M_{\rm ADM}=1.64M_{\odot}$ and
equation of state parameter $\kappa=123.6$. Nevertheless, the curves
remain very close to each other in shape, indicating that the method
to impart NS rotation~\cite{Tichy:2011gw} performs similarly for mixed
BH-NS binaries and for NS-NS binaries~\cite{Tacik:2015tja}.

To investigate numerical convergence of the initial-data sets
presented in figure~\ref{fig:ChiVOmega}, we plot in
figure~\ref{fig:OmegaSeqHamMom} the Hamiltonian and momentum
constraints for a subset of the generated initial data sets, with
$\chi\sim{0.1,0.3,0.5,0.7}$. In general we find the expected
exponential convergence, but there are a few features worth discussing
in the data. The increase in the constraints between the lowest and
second-lowest resolution ($k=0$ vs. $k=1$) arises because the spin is
only activated at the second-lowest resolution, cf. step~\ref{it:1}.
This jump in constraints  monotonically increases with the spin-parameter $\omega$, as we might expect, because the solver has a more difficult task
in adjusting to the abrupt activation of a larger spin. We also note that
at high resolution, in the $\chi\sim 0.7$ curve, we lose exponential
convergence and the curves flatten out around $10^{-6}$. 
This is likely a sign that
the accuracy of the solver is becoming limited, likely by
approximations that go into the solver. $\chi\sim 0.7$ is around the
maximum theoretical neutron star spin, so such difficulties are
expected.

\begin{figure}
\centerline{\hspace*{1.4cm}\includegraphics[scale=0.75]{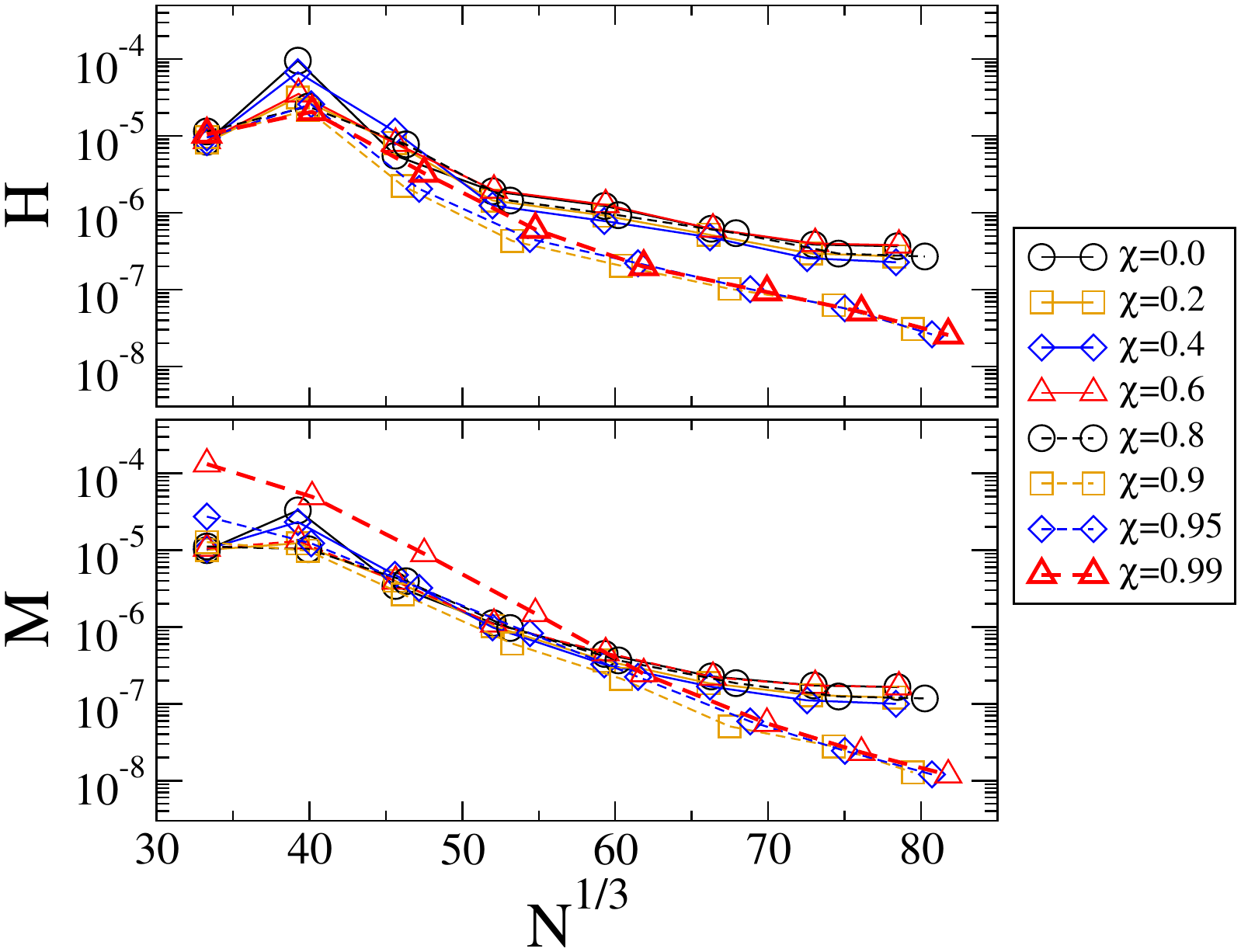}}
\caption[Hamiltonian and momentum for the sequence in $\chi_{\rm
    BH}$.]{\label{fig:chiSeqHam}Hamiltonian constraint (top panel) and momentum
  constraint (bottom panel) versus resolution for our sequence of
  binaries where the black-hole spin is varied from $\chi_{\rm BH}=0$
  to $\chi_{\rm BH}=0.99$. The NS spin parameter is kept constant at $\vec\omega_{\rm NS}=0.017\hat z$ and the mass ratio is $q=7$. 

}
\end{figure}

Continuing the exploration of parameter space, we next vary the black
hole spin $\chi_{\rm BH}$. In partiuclar, we vary the black hole spin
from $\chi_{\rm BH}=0$ to $\chi_{\rm BH}=0.99$, keeping it aligned
with the orbital angular momentum. The other binary parameters are
kept the same as in the ${\tt R14i0\Uparrow}$ initial data set,
specifically $\vec{\omega}_{\rm NS}=0.017\hat z$ and $q=7$. In
Figs.~\ref{fig:chiSeqHam} we plot the Hamiltonian and momentum
constraints, respectively, for this sequence. We find exponential
convergence in all cases. It is interesting to note that the
constraints seem to be lowest at the highest black hole spins,
$\chi_{\rm BH}=0.95$ and $\chi_{\rm BH}=0.99$, while one might expect
these to be the most challenging cases.

\begin{figure}
\centerline{\includegraphics[scale=0.7]{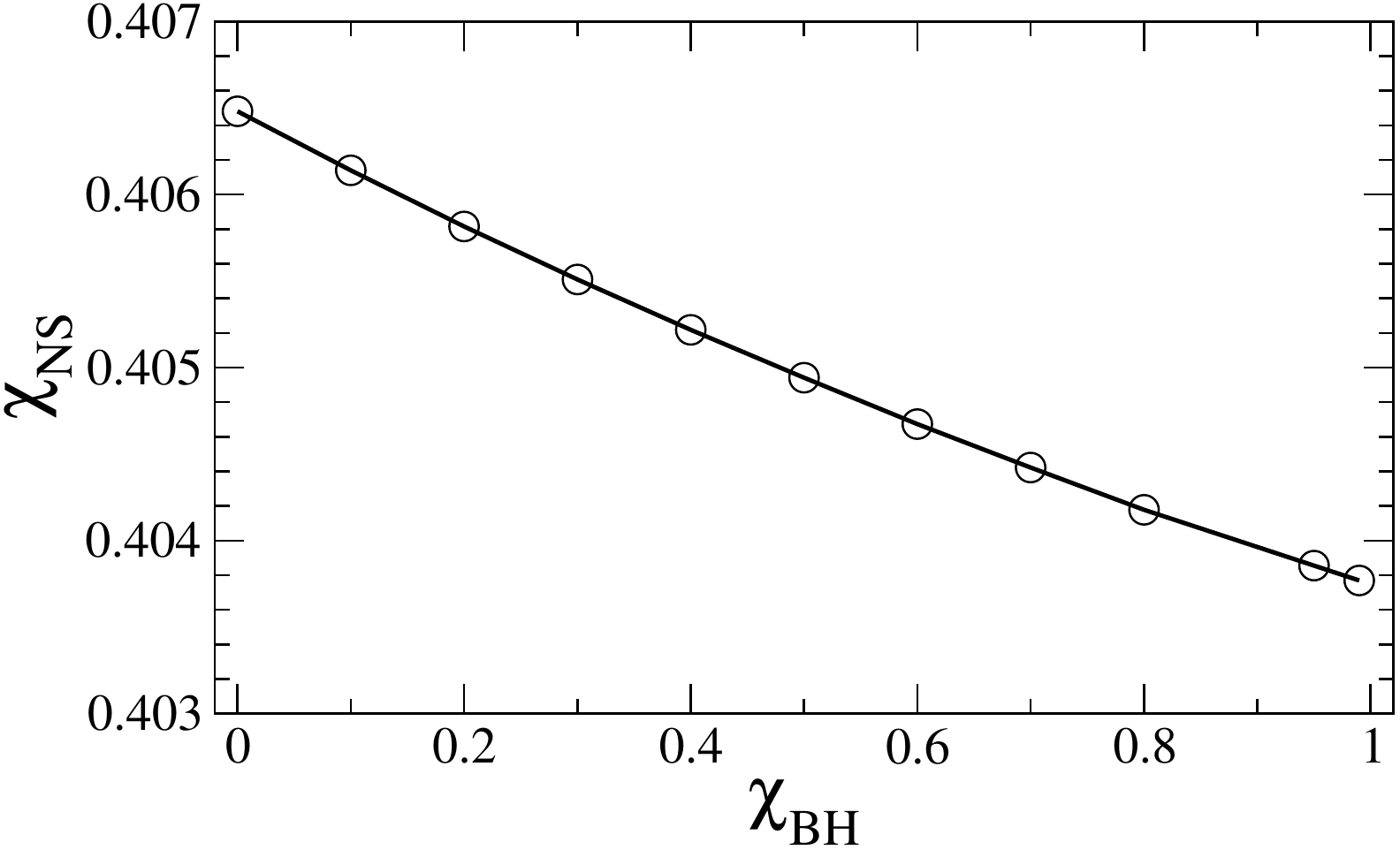}}
\caption[Measured neutron star spin plotted as a function of black
  hole spin.]{\label{fig:chichi}Neutron star spin $\chi_{\rm NS}$ as a
  function of black hole spin $\chi_{\rm BH}$ for this sequence. We
  notice a small downward linear trend.}
\end{figure}

Since this work focuses on neutron star spin, it is interesting to
consider how the measured neutron star spin, $\chi_{\rm NS}$ couples
to other binary parameters. To lowest order, it should depend only on
$\omega_{\rm NS}$, but in practice it may also depend on the
parameters of the black hole or of the orbit. For the sequence of
initial data sets of varying $\chi_{\rm BH}$, figure~\ref{fig:chichi}
presents the neutron star spin $\chi_{\rm NS}$ as a function of
$\chi_{\rm BH}$. $\chi_{\rm NS}$ is nearly constant, dropping by less
than 1\% between $\chi_{\rm BH}=0$ and $\chi_{\rm BH}=0.99$,
confirming that the spin specification for the neutron star almost
completely decouples from the BH spin.

Finally, we consider a sequence of initial-data sets that varies the
mass ratio from $q=2$ to $q=10$. In this sequence we keep the other
binary parameters the same as in the ${\tt R14i0\Uparrow}$ initial
data set and we keep the orbital parameters $M\Omega$ and $D/M$
constant.

As the mass-ratio changes, we expect that the orbital frequency needed
to achieve low eccentricity will also somewhat change. We do not
model this effect, but rather keep all other binary parameters are the
same as in the {\tt R14i0$\Uparrow$} run. In particular, the orbital
parameters $M\Omega$ and $D/M$ are constant. While not the most
accurate way of choosing these parameters, as it is only correct to
Newtonian order, it suffices for the present purpose of testing
robustness of the initial-data solver.

To estimate the impact on the eccentricity of the constructed
initial-data sets, we use the post-Newtonian expansion of the orbital
frequency of a BBH in a \emph{circular} orbit (Eq. 228 of
\cite{Blanchet2006}):
\begin{equation}
\Omega^2=\frac{GM}{r^3}\left(1+(-3+\nu)\gamma+\left(6+\frac{41}{4}\nu+\nu^2\right)\gamma^2+...\right).
\end{equation}
Here $\nu=m_1m_2/(m_1+m_2^2)=q/(1+q)^2$ is the symmetric mass ratio, and
$\gamma=GM/Dc^2$. Keeping $D$ and $M$ constant, the quantity $M\Omega$
varies by approximately 3\% in the mass ratio range we consider.
Therefore, we expect that the eccentricity of our initial-data sets varies
by only a few percent between $q=2$ and $q=10$.

\begin{figure}
\centerline{\hspace*{1.5cm}\includegraphics[scale=0.75]{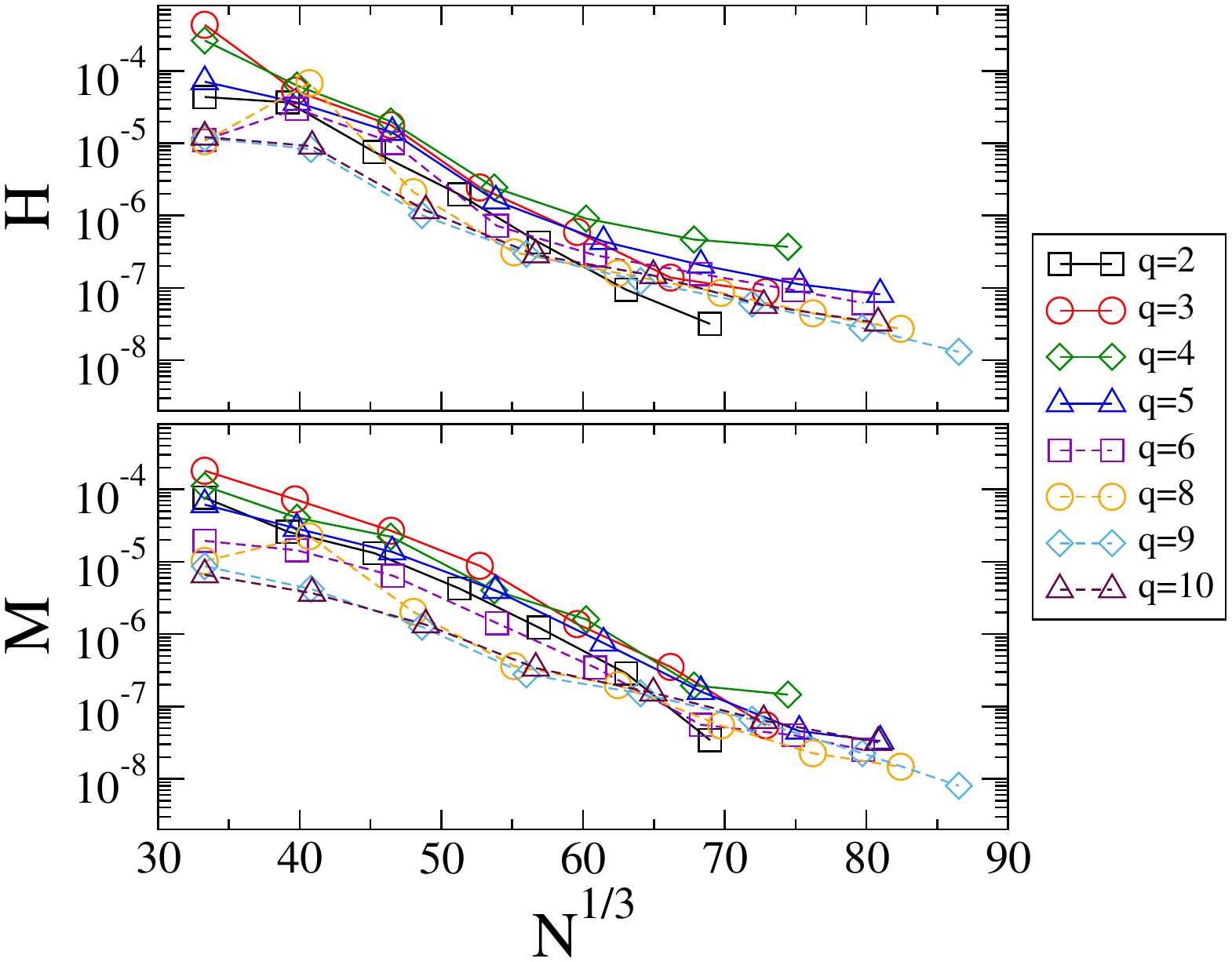}}
\caption[Hamiltonian and momentum constraint for the $q$
sequence.]{\label{fig:qSeqHam}Hamiltonian constraint (top panel) and
  momentum constraint (bottom panel) versus resolution for our
  sequence of binaries where the mass ratio is varied from $q=2$ up to
  $q=10$. The NS spin parameter is kept constant at
  $\vec\omega_{\rm NS}=0.017\hat z$ and the black hole spin is
  $\chi_{\rm BH}=0.9$. }
\end{figure}

To assess convergence, we plot the Hamiltonian and momentum
constraints for this sequence in Fig.~\ref{fig:qSeqHam}. We find
exponential convergence in all cases. Interestingly, no clear pattern
in $q$-space emerges.

\begin{figure}
\centerline{\includegraphics[scale=0.7]{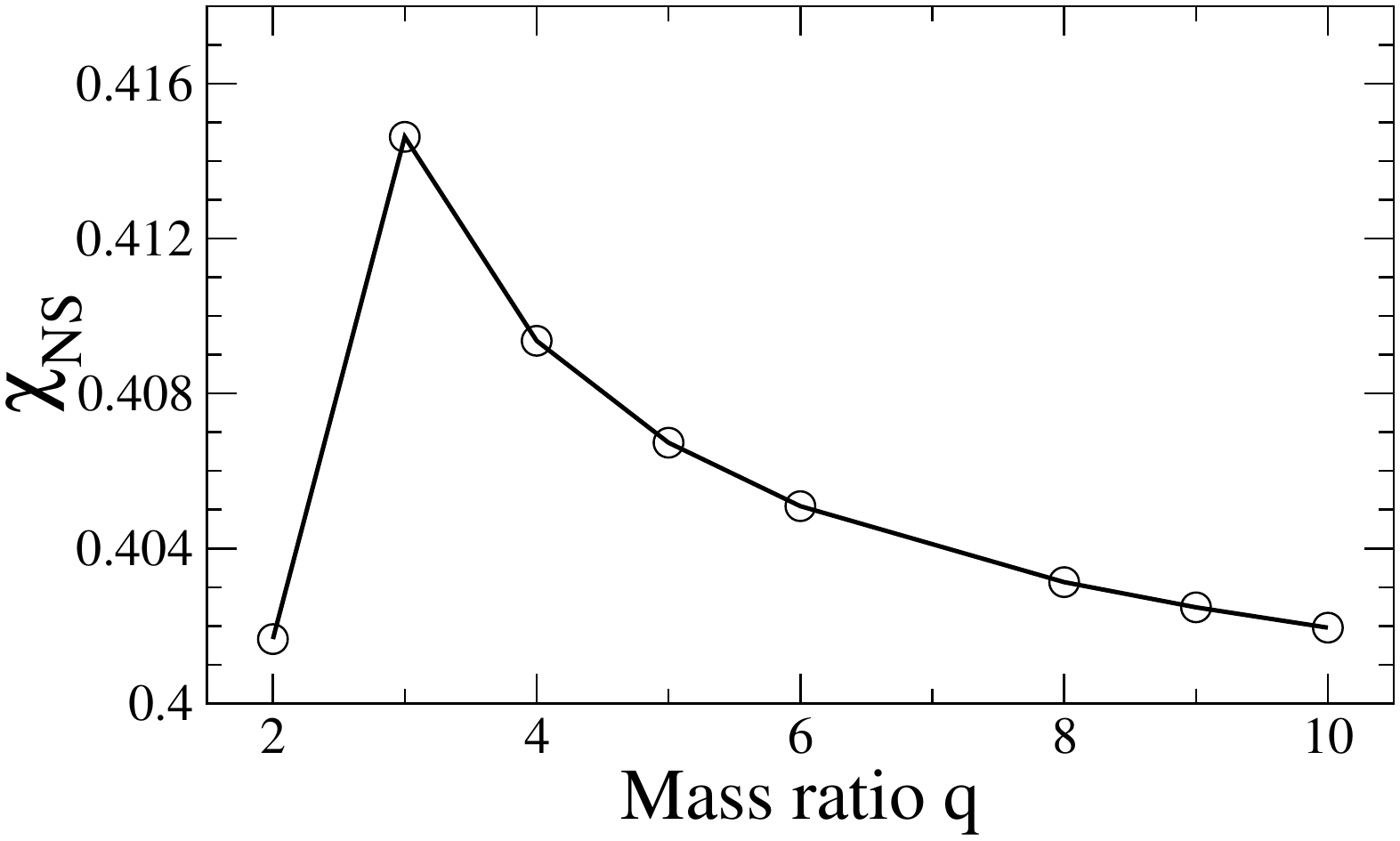}}
\caption[$\chi_{\rm NS}$ as a function of mass ratio
$q$.]{\label{fig:qchi}Neutron star spin $\chi_{\rm
    NS}$ as a function of mass ratio $q$ for this sequence. We notice
  a small downward trend for $q \geq 3$.}
\end{figure}

Figure~\ref{fig:qchi} plots the neutron star spin as a function of mass-ratio.
Having kept $\vec\omega_{\rm NS}$ constant across this sequence, 
we indeed find that the physical NS spin is approximately constant, too, varying less than $2$ percent.  Although there is not a great amount of variation,
apart from $q=2$, there is a clear trend of $\chi_{\rm NS}$ decreasing
with $q$. Again, however, the effect is quite small and we do not seek
to explain it.

\section{Conclusion}
\label{sec:Conc}

In compact object binaries containing neutron stars, the spin of the
neutron star(s) forms part of the parameter space of such binaries.
In order to constrain neutron star spin \emph{directly from
  gravitational wave observations}, one must know the impact of the
neutron star spin on the evolution of the compact object binary,
i.e. on the emitted waveforms and on the electro-magnetic signature.

This paper lays foundations for such studies by constructing
initial-data sets of BH-NS binaries with arbitrary neutron star spins.
To our knowledge, this is the first time initial data has been created
for BH-NS binaries with spinning neutron stars. To impart spin on the
neutron star, we carry over the formalism developed
by~\cite{Tichy:2011gw} and used
in~\cite{Bernuzzi:2013rza,Tacik:2015tja,Dietrich:2015pxa} to create
initial data for NS-NS systems with arbitrary spins.

Two new numerical tricks were found to be necessary to get convergent
initial data - setting a maximum radius out to which to apply
$W^i=\epsilon^{ijk}\omega^jr^k$, and only activating the neutron
star spin after the first AMR-iteration of the initial data solver has
completed. We create initial data sets across a large
portion of the BH-NS binary parameter space. 

First, we present a comprehensive study of initial-data sets with various NS spins, restricting to $q=7$ and $\chi_{\rm BH}=0.9$.
This first study spans three different equations of state (all
$\Gamma=2$ polytropes), different neutron star spin magnitudes,
different neutron star spin orientations, and four different black
hole spin orientations. Subsequently, we construct initial data with
spinning NS for mass-ratios from $q=2$ to $q=10$, and for black-hole
spins $0\le \chi_{\rm BH}\le 0.99$, the latter well exceeding the
standard Bowen-York limit on black hole
spin~\cite{Lovelace2008,DainEtAl:2008}. Finally, we explore the range
of possible NS spin magnitudes, and find that the presented numerical
techniques can successfully construct initial data with neutron star
spins ranging from $\chi_{\rm NS}=0$ to $\chi_{\rm NS}\sim0.7$ (near
the maximum theoretical spin for neutron stars).

Future research will involve running evolutions of these, or similar,
initial data sets. Some of the 36 initial data sets of our first study
(Table~\ref{tab:36Sets}) can be used to investigate how neutron star
spin affects tidal disruption of the star by the black hole, and how
it affects the disk that is formed. The orbital phase evolution can
also be examined and compared to Post-Newtonian\cite{Blanchet2014} or
other analytic predictions such as Effective-One-Body
(EOB)~\cite{Buonanno99,Taracchini:2013rva,Pan:2013rra}.

One can also explore the maximum mass of accretion disks and ejecta as
a function of NS spin. Ref.~\cite{Lovelace:2013vma} finds a very large
disk with a black hole spin of $\chi=0.97$ and mass ratio $q=3$. Keeping these BH and NS parameters, but adding spin on the neutron star will cause the NS' material to be less strongly bound and may increase the disk mass even further.

\ack

We gratefully acknowledge
support for this research at Cornell and Caltech from the
Sherman Fairchild Foundation and NSF grants PHY-1306125, PHY-1404569,
AST-1333520, and AST-1333129, and at CITA from NSERC of Canada,
the Canada Research Chairs Program, and the Canadian Institute for
Advanced Research. Calculations were performed at the GPC supercomputer at
the SciNet HPC Consortium~\cite{scinet}; SciNet is funded by: the
Canada Foundation for Innovation (CFI) under the auspices of Compute
Canada; the Government of Ontario; Ontario Research Fund (ORF) --
Research Excellence; and the University of Toronto.
\\
\\

\bibliographystyle{iopart-num}
\bibliography{../References/References}

\end{document}